\documentclass[final,twocolumn,journal]{IEEEtran}
\usepackage{graphicx}      
\usepackage{verbatim}
\usepackage{amsmath}
\usepackage{amssymb}
\usepackage{float}
\usepackage{color}
\usepackage{epstopdf}
\usepackage{yhmath}
\usepackage[caption=false]{subfig}
\usepackage{amsmath}

\renewcommand{\t}{^{\mbox{\tiny {T}}}}

\newcommand{\eproof}{\hfill\rule{2mm}{2mm}}

\newcommand{\bstate}{\medskip\begin{state} }
\newcommand{\estate}{ \hfill  \rule{1mm}{2mm}\medskip\end{state}}

\newcommand{\bass}{\medskip\begin{ass} }
\newcommand{\eass}{ \hfill  \rule{1mm}{2mm}\medskip\end{ass}}

\newcommand{\brem}{\medskip \begin{remark}  }
\newcommand{\erem}{\hfill \rule{1mm}{2mm}\medskip
\end{remark} }
\newcommand{\bthm}{\medskip\begin{theorem}  }
\newcommand{\ethm}{ \hfill  \rule{1mm}{2mm} \medskip
\end{theorem} }
\newcommand{\blem}{\medskip\begin{lemma}  }
\newcommand{\elem}{ \hfill \rule{1mm}{2mm}\medskip
\end{lemma} }
\newcommand{\bcorollary}{\medskip\begin{corollary}  }
\newcommand{\ecorollary}{  \hfill \rule{1mm}{2mm}\medskip
\end{corollary} }
\newcommand{\bdefn}{\medskip\begin{definition}}
\newcommand{\edefn}{  \hfill \rule{1mm}{2mm}\medskip
\end{definition} }
\newcommand{\bproposition}{\medskip\begin{proposition} }
\newcommand{\eproposition}{\hfill \rule{1mm}{2mm}\medskip
\end{proposition} }
\newcommand{\bexample}{\medskip\begin{example} \rm}
\newcommand{\eexample}{ \hfill \rule{1mm}{2mm}\medskip
\end{example} }
\newcommand{\proofnow}{\noindent{\bf Proof: }}
\newcommand{\commadef}{   
 \def\OldComma{,}
    \catcode`\,=13
    \def,{%
      \ifmmode%
        \OldComma\discretionary{}{}{}%
      \else%
        \OldComma%
      \fi%
    }%
 }

\newtheorem{theorem}{\bf Theorem}[section]
\newtheorem{ass}{\bf Assumption}[section]
\newtheorem{lemma}{\bf Lemma}[section]
\newtheorem{definition}{\bf Definition}[section]
\newtheorem{remark}{\bf Remark}[section]
\newtheorem{corollary}{\bf Corollary}[section]
\newtheorem{proposition}{\bf Proposition}[section]
\newtheorem{example}{\bf Example}[section]
\newtheorem{state}{\bf Assumption}[section]

\begin{document}

\title{Synchronization  of Power Systems  and Kuramoto Oscillators: A Regional Stability Framework}

\author{Lijun Zhu, and David J. Hill \IEEEmembership{Life Fellow, IEEE}
\thanks{This work was supported by The University of Hong Kong Research Committee Post-doctoral Fellow Scheme. }
\thanks{Lijun Zhu and David J. Hill are 
with Department of Electrical and Electronic Engineering, The University of Hong Kong, Hong Kong. (e-mail: {ljzhu,dhill}@eee.hku.hk).}
}

\maketitle
\begin{abstract}
The transient stability of power systems and   synchronization of non-uniform Kuramoto oscillators are closely related problems. In this paper, we develop a novel regional stability analysis framework based on the proposed region-parametrized Lyapunov function to solve the problems. Also,  a new synchronization definition  is introduced and characterized by frequency boundedness and angle cohesiveness, the latter of which requires angles of any two connected nodes rather than any two arbitrary nodes to stay cohesive. It allows to take  power fluctuations into explicit account as disturbances and  can lead to  less conservative stability condition. 
Applying the analysis framework, we derive two algebraic stability conditions for power systems that relate the underlying network topology and system parameters  to the stability. Finally, to authors' best knowledge, we first explicitly give the estimation of region of attraction for power systems. The analysis is verified via numerical simulation showing that  two  stability conditions can complement each other for predicting the stability. 
\end{abstract}



\section{Introduction}
Power systems are a class of heterogeneous complex networks composed of load and generator buses connected via electric lines.
The angle  stability of power systems refers to the ability of bus angles to stay synchronism  after severe faults or when the system experiences  power fluctuations. It ensures stable and secure system operation  to deliver electric power reliably from generators to loads.
Small-disturbance and transient stability analysis are two classes of stability analysis.
Small-disturbance stability concerns the stability issues of power systems under disturbances of small scale and usually uses the eigenvalue-based method following the model linearization.
Transient stability considers the stability under rather large disturbances and the stability result is effective in a larger region of interest than the small-disturbance stability. 

Transient stability assessment approaches  are categorized into direct time-domain simulation and energy function methods.
Time-domain simulation  assesses the stability with respect to a given
fault or disturbance by means of numerical simulation \cite{Huang2012,Nagel2013}.  On the contrary, the energy
function method adopts  Lyapunov stability theory and relies on a class of energy functions to determine the system stability. It identifies  critical unstable equilibrium points (UEPs)
\cite{El-Abiad1966} such as closest UEP or controlling UEP \cite{Chiang1995,Chang1995} which are used to infer the stability. For instance, when the post-fault energy is less than the energy of closest UEP,   the system trajectory is guaranteed to converge towards the system equilibrium.  Time-domain simulation
is less intuitive  and requires intensive computation
especially for large-scale power systems but guarantees  the accuracy   if the precise
modeling of the system is available \cite{Majumder2013,Miao2011}. In comparison,  energy
function   method provides more insights and is less computing intensive, although the estimated region of attraction is conservative. 

The transient stability of power systems is also closely related to the synchronization of celebrated Kuramoto oscillators in terms of  dynamic model and phase (angle) behavior. For  conventional power systems, the dynamic model of synchronous generators under the over-damped assumption   can be approximated by the modified Kuramoto model \cite{Dorfler2012}.  For microgrids,  the droop-controlled frequency dynamics of the inverter-interfaced energy sources  resemble  Kuramoto model \cite{Ainsworth2013,Simpson-Porco2012,Simpson-Porco2013}. However, the network structures of Kuramoto oscillators and power networks sometimes are different. The complete graph structure  is usually assumed for Kuramoto model and facilitates it to study necessary and sufficient synchronization conditions, while the network structure of power systems is usually irregular. In the early work, a network reduction method called Kron reduction (\cite{Kundur1994}) was introduced to simplify the network. For instance, reference \cite{Athay1979,Lueders1971} considered  loads were modeled as constant impedances and used Kron reduction  to absorb loads into lines and reduce the original meshed power network into a network of generators. 
The Kron reduction simplifies the power network but has two drawbacks:
the loss of the original topological
information and inclusion of higher transfer conductances resulting from load absorption. 
The former makes it difficult to explore  relation between stability and the original network topology, while
the latter makes an unsolved problem  to develop general Lyapunov functions.
Later,  Bergen and Hill \cite{Bergen1981}
proposed the network-preserving model of  conventional power systems with frequency-dependent loads for which Lyapunov functions in Lur'e-Postnikov
form (\cite{Bergen1981,Hill1982,Hill1989}) were proposed. The network-preserving model allows for    more precise dynamic modeling of  loads, while  the original network structure is retained.


The synchronization of Kuramoto oscillators refers to phase synchronization if natural frequencies of oscillators  are identical or  phase locking otherwise, i.e., phases of oscillators are distributed in a  pattern. The   phase locking coincides with transient stability definition of power systems.  The research on Kuramoto oscillators mainly focus on finding  necessary
 (see, e.g., \cite{Chopra2009,Verwoerd2009,Jadbabaie2004}) and sufficient synchronization
conditions (see, e.g.,\cite{Chopra2009,Doerfler2011,Franci2010,Chung2010}). The work \cite{Doerfler2011,Dorfler2012} first  linked  the stability  of network-reduced power systems with synchronization of Kuramoto oscillators, and adopted notations such as  phase cohesiveness and frequency synchronization to  characterize the stability for power systems. Motivated by \cite{Hill2006}, \cite{Dorfler2012} also showed that the network topology has a crucial impact on the stability of power systems.  

Over the last decade, the increasing integration of renewable energy into power grids has
been motivated by environmental and economic benefits  and
continues as the enabling technology innovation progresses. Microgrid
is one of  promising technologies that can integrate large amount of 
renewable energy such as  solar, wind power and geothermal systems, and fulfills  the potential of distributed generation  (\cite{Lasseter2002,Piagi2006}) in a systematic way.   In general,  energy sources are fed    via power-electronic converters, whose characteristics are determined by the internal control logic and are largely different from the conventional synchronous machine based power generators \cite{Simpson-Porco2012,Simpson-Porco2013,Schiffer2014,Schiffer2016}.   
The power generation of renewable energy is  intermittent, stochastic and  subjected
to weather condition. On the other hand,  the demand-side
activities become complicated and less predictable. The generation of renewable energy and complicated load activities may cause 
fluctuations in power
systems, which  have not been accounted for in the existing stability analysis.

In this paper, we will show that the transient stability of power systems  is related to the synchronization of  non-uniform Kuramoto oscillators.  
The objective of this paper is  to establish a general analysis framework based on energy (Lyapunov) functions to study  the transient stability for power systems  and the synchronization of non-uniform Kuramoto oscillators. The main  contributions are summarized as follows. First, we introduce a new definition for
the synchronization of power systems and   Kuramoto oscillators characterized by angle differences across any physical lines being less than $\pi$,  which complements the definition of phase cohesiveness in \cite{Dorfler2012} that considers angle differences of any arbitrary two angles in the system. Second, with the recognition that renewable energy  has stochastic and intermittent nature, we explicitly consider  the energy fluctuations as  disturbances to power systems  and analyze their impact on stability. Third, we propose a general stability analysis method based on region-parametrized Lyapunov function whose bounds are parametrized by the size of  region of interest. The stability analysis gives the  existence condition of positively invariant sets in terms of the energy and boundedness in terms of the state which can be used to obtain the condition  for angle cohesiveness and frequency boundedness. Fourth, applying the stability analysis framework, we derive two algebraic conditions for power systems in terms of the new definition and a definition similar to that in \cite{Dorfler2012} that both relate the underlying network topology and system parameters  to the stability. Finally, to authors' best knowledge, we first explicitly give the estimation of  the region of attraction for power systems.
This paper is a strengthened extension to  our conference paper \cite{Zhu2016a} in several aspects including the aforementioned third, fourth and last points. In addition, we will explain the motivation of the new synchronization definition  using an example and show positively invariant set from bus angle perspective in addition to energy perspective in \cite{Zhu2016a}.  Also, we derive additional stability condition in Theorem \ref{thm:robust_c}.


The rest of the paper is structured as follows. Section \ref{sec:PF}
presents the structure-preserving model of  power systems and introduces
the first stability definition and describes the problem to be studied.
In Section \ref{sec:syn}, we introduce a new stability definition using a motivating example and present a coordinate transformation.
In Section \ref{sec:SA}, we propose  stability analysis framework based on region-parametrized Lyapunov functions and apply it to obtain two stability conditions for power systems in Section \ref{sec:ST}.
Section \ref{sec:ext} extends the stability analysis in Section \ref{sec:ST} to non-disturbance scenario and explicitly gives the estimation of region of attraction for power systems.
 Section \ref{sec:sim} verifies  theoretical results
on the IEEE 9-bus test system using numerical simulation. The paper
is concluded in Section \ref{sec:Conclusion}.

\textbf{\textit{Notations.}} For a scalar $x\in\mathbb{R}$, $\mbox{sinc}(x)=\sin(x)/x$.
For a vector $x=[x_{1},\cdots,x_{n}]\t\in\mathbb{R}^{n}$, $\|x\|$
and $\|x\|_{\infty}$ are the 2-norm and the $\infty$-norm of vector
$x$ and $\sin(x):=[\sin(x_{1}),\cdots,\sin(x_{n})]\t$. The vector
$e_{n}$ is a column vector of dimension $n$ with all elements being
1. The notations from algebraic graph theory is defined as follows. An undirected $\mathcal{G}=(\mathcal{V},\mathcal{E})$
consists of a set of vertices $\mathcal{V}=\{1,\cdots,n\}$ and a
set of undirected edges $\mathcal{E}\subseteq\mathcal{V}\times\mathcal{V}$. An undirected
edge of $\mathcal{E}$ from node $i$ to node $j$ is denoted by $(i,j)$,
meaning that nodes $\mathcal{V}_{i}$ and $\mathcal{V}_{j}$ are interconnected
with each other. The edge weight is denoted by $a_{ij}$ where $a_{ii}=0$ and $a_{ij}=a_{ji}>0$
for $(j,i)\in\mathcal{E}$. The Laplacian of the graph $\mathcal{G}$
is denoted by $L=[l_{ij}]\in\mathbb{R}^{n\times n},$ where $l_{ii}=\sum_{j=1}^{n}a_{ij}$
and $l_{ij}=-a_{ij}$ if $i\neq j$. Denote by $\mathcal{E}_{k}$ the
$k$th edge of $\mathcal{E}$ where $k\in\{1,\cdots,|\mathcal{E}|\}$,
$|\mathcal{E}|$ the number of edges, and $B\in\mathbb{R}^{n\times|\mathcal{E}|}$
the incidence matrix whose component is $B_{ik}=1$ if node $i$ is
the sink node of edge $\mathcal{E}_{k}$, $B_{ik}=-1$ if it is
the source node and $B_{ik}=0$ otherwise. As a result, one can have $L=BA_{v}B\t$ where $A_{v}=\mbox{diag}(\{a_{ij}\}_{(i,j)\in\mathcal{E}})$
is the diagonal matrix with diagonal elements being edge weights. $\mathcal{G}_{c}$ is called the complete graph induced by $\mathcal{G}$,
if $\mathcal{G}_{c}=(\mathcal{V},\mathcal{E}_{c})$ is an undirected
complete graph with the same set of nodes as $\mathcal{G}$, for which
$B_{c}$ is the incidence matrix.

\section{System Model and Problem Formulation} \label{sec:PF}
In this paper, we study the  first-order dynamics  
\begin{equation}
d_{i}\dot{\theta}_{i}=p_{i}(t)-\sum_{j=1}^{n}a_{ij}\sin(\theta_{i}-\theta_{j}),\;i=1,\cdots,n.\label{eq:FO}
\end{equation}
The model (\ref{eq:FO}) can represent Kuramoto oscillators, conventional power systems with over-damped  synchronous generators \cite{Dorfler2012}, lossy \cite{Schiffer2014}  and lossless \cite{Ainsworth2013,Simpson-Porco2012} microgrids.
For instance, the  network-preserving model of  lossless microgrids with inverter-based energy sources and loads can be described by (\ref{eq:FO}) in which $\theta_{i}$ is the phase angle of the voltage $V_i$  at bus $i$. The network parameter is $a_{ij}=|V_i||V_j||B_{ij}|$ where $B_{ij}$  is the susceptance of the line connecting bus $i$ and $j$,  $|V_i|$ and $|V_j|$ are magnitudes of voltage $V_i$ and $V_j$, respectively.   Note that  $a_{ij}>0$ if two buses are connected, and $a_{ij}=0$ otherwise.
The net power injected from the network  $p_{e,i}=-\sum_{j=1}^n a_{ij}\sin(\theta_{i}-\theta_{j})$.
 Let $\mathcal{V}=\mathcal{V}_l \cup \mathcal{V}_s$ where $\mathcal{V}_l={1,\cdots,l}$ and $\mathcal{V}_s={l+1,\cdots,n}$ are  index sets for the load   and energy source buses, respectively. For $i\in\mathcal{V}_l$, the equation (\ref{eq:FO}) describes the power balance between power injection and power consumed by the load \cite{Ainsworth2013}, for which we adopt the frequency-dependent load  (\cite{Bergen1981}) where $p_i<0$ is the nominal consumption and $d_i$ is the frequency-dependent parameter. For $i\in\mathcal{V}_s$,  energy sources  are equipped with   AC-AC or DC-AC inverter and their  dynamics  are determined by the internal control logic of the inverters which normally implement droop control
\cite{Ainsworth2013} or maximum power point tracking (MPPT) \cite{Doerfler2016}. For either control strategy,  the equation (\ref{eq:FO}) depicts the
power balance between energy consumption by internal load, power supply by energy sources and power delivery to microgrids.
For droop control, $d_{i}$
and $p_{i}$ are related to  parameters and  setpoints of the droop
control (see \cite{Ainsworth2013}), while for MPPT control, $p_{i}$ is the maximum power output and
$d_{i}$ is related to the internal frequency-dependent load.  $p_i$ can be simply  regarded as the power supplied  by  $i$th energy source.

\brem In contrast, the classic Kuramoto oscillators are 
\begin{equation*}
\dot{\theta}_{i}=p_{i}-\frac{K}{n}\sum_{j=1}^{n}\sin(\theta_{i}-\theta_{j}),\;i=1,\cdots,n.
\end{equation*}
where $p_{i}$ is the natural frequency of $i$th oscillator, $K$ is the coupling strength and the  network graph has all-to-all connections.  The model (\ref{eq:FO}) is also called non-uniform Kuramoto oscillators that was studied in \cite{Dorfler2012}. Because the model  (\ref{eq:FO})  has non-complete interconnection and non-uniform coefficient $d_i$, it is more challenging to study the synchronization.  \erem
\brem The model of lossy microgrids can also be written in the form of  (\ref{eq:FO}), with $p_i(t)$ replaced by $p_i'(t)$, as 
$
d_{i}\dot{\theta}_{i}=p_{i}'(t)-\sum_{j=1}^{n}a_{ij}\sin(\theta_{i}-\theta_{j}),\;i=1,\cdots,n,\label{eq:lossy_FO}
$
where $p_{i}'(t)=p_{i}(t)-|V_i|^2G_{ii}+\sum_{j=1}^{n}|V_i||V_j|G_{ij}\cos(\theta_{i}-\theta_{j})$. The second and last term in $p_{i}'(t)$ are the power transfer induced by non-zero conductances $G_{ij}$.  This model   can also represent the network-reduced model of conventional power systems with over-damped  synchronous generators \cite{Dorfler2012}. \erem

The term $p_i$ is normally assumed to be constant in the literature of power systems and Kuramoto oscillators, however it is worth mentioning that
$p_i$ in this paper might be time-varying due to load and renewable generation fluctuations. For instance, for MPPT control the maximum power outputs of the renewable energy such as PV and wind power normally vary with the weather condition. 

The dynamical system (\ref{eq:FO}) can be put in a
vector form, with $\theta=[\theta_1,\cdots,\theta_n]\t$,  as follows
\begin{equation}
\dot{\theta}=-D^{-1}\left(BA_{v}\sin(B\t\theta)-p_{f}(t)\right)\label{eq:FOVO}
\end{equation} 
where $B$ is the incidence matrix of the power network $\mathcal{G}$,  $D=\mbox{diag}(d_{1},\cdots,d_{n})$ is the coefficient matrix, $p_{f}=[p_{1},\cdots,p_{n}]\t\in\mathbb{R}^{n}$  is called 
power profile vector. Define 
\begin{equation}
\theta_c=B_c\t\theta, \label{eq:tc}
\end{equation} where $B_c$ is the incidence matrix of the induced complete graph $\mathcal{G}_c$. Hence, the elements in $\theta_{c}$ are ${\theta}_{i}-{\theta}_{j}$
for $i\neq j,\forall i,j\in\{1,\cdots,n\}$. 
The stability in terms of synchronization for (\ref{eq:FOVO})  with notations of phase cohesiveness and frequency synchronization was introduced in \cite{Dorfler2012}, adapted in \cite{Zhu2016} and revised as follows. A few notations are adopted from \cite{Dorfler2012} for the purpose of self-containedness.   
The torus is the set $\mathbb{T}^{1}=[0,2\pi]$ where $0$ and $2\pi$
are associated with each other. An angle is a point $\theta\in\mathbb{T}^{1}$
and an arc is a connected subset of $\mathbb{T}^{1}$. The $n$-torus
is the Cartesian product $\mathbb{T}^{n}=\mathbb{T}^{1}\times\cdots\times\mathbb{T}^{1}$.

\bdefn \label{def:cohesive_c}(Phase Cohesiveness
and Frequency Boundedness). A solution $\theta(t):\mathbb{R}^{+}\rightarrow\mathbb{T}^{n}$
is then said to be phase cohesive if there exists a $\gamma\in[0,\pi)$
such that $\|\theta_c\|_\infty \leq \gamma$. A solution $\dot{\theta}(t):\mathbb{R}^{+}\rightarrow\mathbb{R}^{n}$
is then said to be frequency bounded if there exists a $\varpi_{o}$
such that $\|\dot{\theta}(t)\|_{\infty}\leq\varpi_{o}$. \edefn

In \cite{Dorfler2012},  the    transient stability of power systems and synchronization of non-uniform Kuramoto oscillators were studied in terms of   phase cohesiveness and frequency synchronization, that is $\lim_{t\rightarrow\infty} \dot \theta (t) = c e_n$ for some constant $c\in\mathbb{R}$. Since $p_i$ is time-varying in this paper, the system is not able to achieve the frequency synchronization   but  rather frequency boundedness in  Definition \ref {def:cohesive_c}. As shown in next section, the phase cohesiveness in  Definition \ref {def:cohesive_c} may lead to some conservativeness and thus we will introduce a new phase cohesiveness definition later. 
The main \textit{objective} of this paper is to investigate the synchronization  of power systems and non-uniform Kuramoto oscillators (\ref{eq:FOVO})  in the sense of Definition \ref{def:cohesive_c} and a new definition to be given in next section.

\section{A New Synchronization Definition And Equilibrium Subspace} \label{sec:syn}
\subsection{A New Synchronization Definition}
The notation of phase cohesiveness in the sense of Definition \ref{def:cohesive_c} was graphically explained in Example
2.2 of \cite{Doerfler2011} for a two-bus system. The following example
uses a three-bus system to complement the explanation in \cite{Doerfler2011},
explains the role of coupling forces between buses and more importantly
motivates a new stability notation. A few more notations are helpful.
For a set of angles $(\theta_{1},\cdots,\theta_{n})$, define $\wideparen{\theta_{1}\cdots\theta_{n}}$
the arc that starts at $\theta_{1}$, ends at $\theta_{n}$ and travels
across angles in the order of $(\theta_{1},\cdots,\theta_{n})$ and
$\mathcal{A}(\wideparen{\theta_{1}\cdots\theta_{n}})$ is its length. 

\begin{figure}[h]
\centering%
\subfloat[Angles $\theta_{A}$, $\theta_{B}$, $\theta_{C}$
in Torus $\mathbb{T}$]{
\includegraphics[clip,scale=0.35]{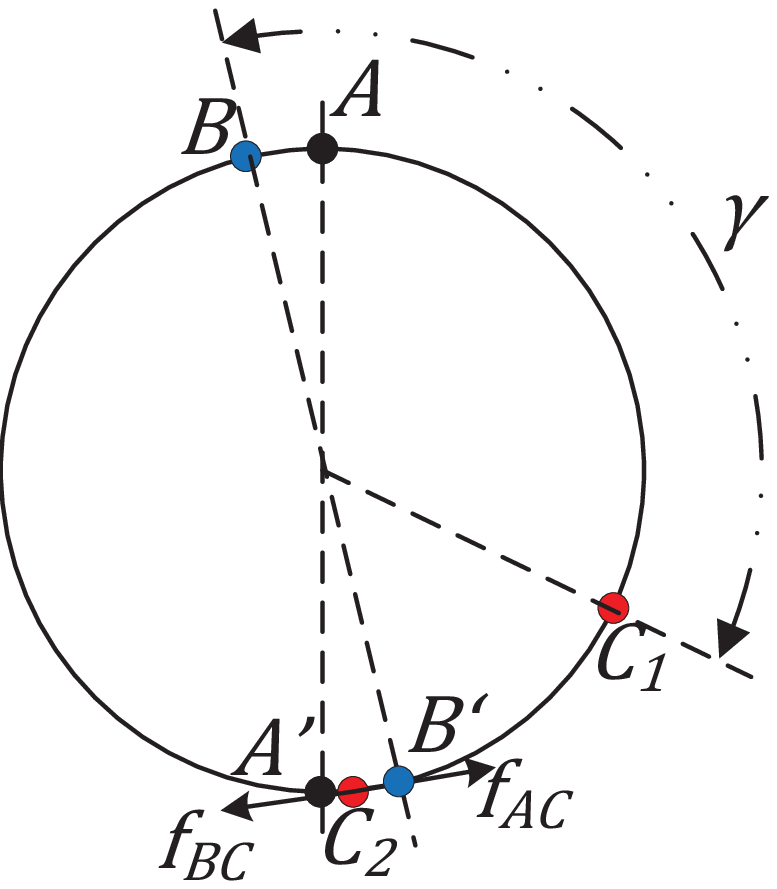}}  \hspace{0.5cm}
\subfloat[ A complete network configuration.]{
\includegraphics[scale=0.35]{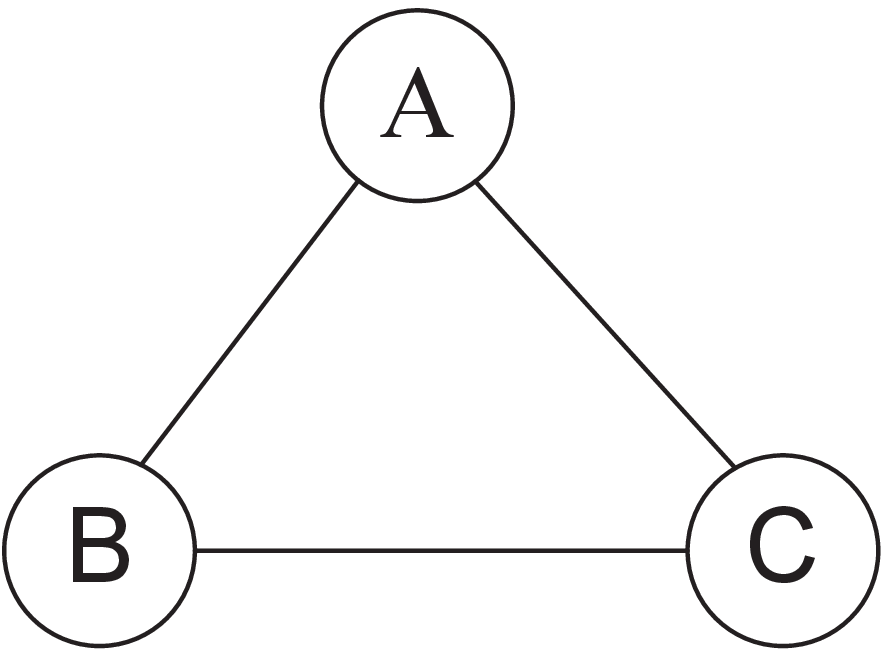}}\hspace{0.5cm}
\subfloat[ A non-complete network configuration.]{
\includegraphics[scale=0.35]{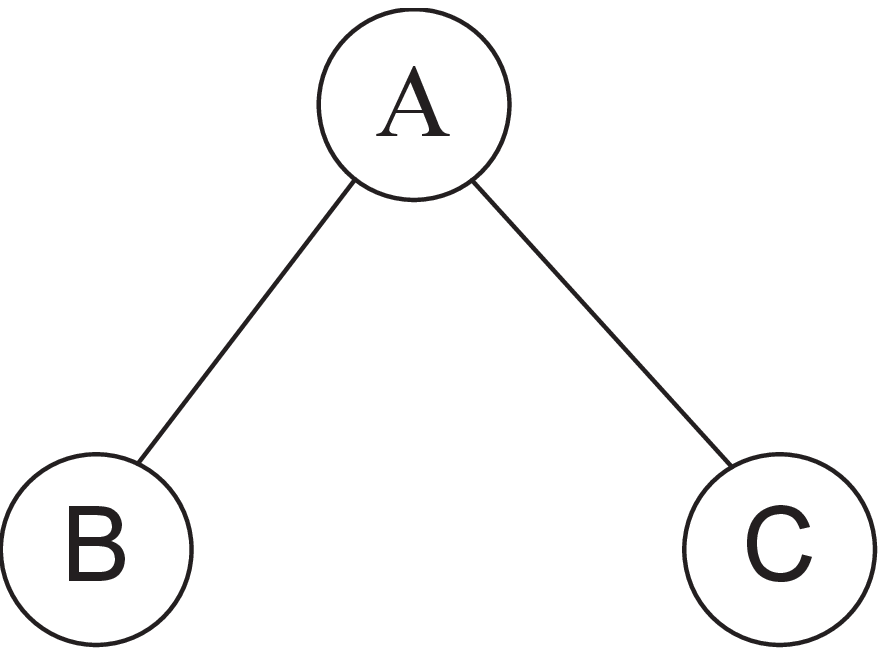}}
\caption{Illustration of phase cohesiveness for the power system in difference
network configurations. }
\label{fig:example}
\end{figure}

\bexample \label{exp:angle} Consider the three-bus system (\ref{eq:FOVO}) with zero power profile $p_{i}=0,$ $i=1,\cdots,3$ and its  network topology is illustrated in Fig.
\ref{fig:example}.b.
The buses are labeled $A$, $B$, $C$ and connected in an all-to-all
fashion. As illustrated in Fig. \ref{fig:example}.a, the bus angle
in a torus is marked as a point in the circle. The desired synchronization
behavior is that all three angles converge to a common value. Suppose,
due to external disturbances, angle $\theta_{C}$ is disturbed to
the position $C_{1}$ within arc $\wideparen{AB'}$. $\wideparen{BAC_{1}}$
is the shortest arc containing all $(\theta_{A},\theta_{B},\theta_{C})$
in its interior and $\mathcal{A}(\wideparen{BAC_{1}})<\pi$. In this
case, the coupling forces among them play an active role of holding
bus angles together. No matter angle $\theta_{C}$ leads ahead/lags
behind angle $\theta_{A}$, it will results in negative/positive coupling
force $-a_{CA}\sin(\theta_{C}-\theta_{A})$ at bus $C$, decelerating/accelerating
the angle $\theta_{C}$ to force these two angles together. This argument
also applies to angle pairs $(A,B)$ and $(B,C)$. As a result, the
length of the arc stays $\mathcal{A}(\wideparen{BAC_{1}})<\pi$, and
then angles are cohesive in the sense of Definition \ref{def:cohesive_c}.
This mechanism is effective if $\mathcal{A}(\wideparen{BAC_{1}})<\pi$
which coincides with the cohesiveness condition $\max_{i,j\in\{A,B,C\}}|\theta_{i}-\theta_{j}|\leq\gamma<\pi$.
However, if $\theta_{c}$ is disturbed farther away beyond the position
$B'$ to the position $C_{2}$ in Fig. \ref{fig:example}.a, say within
$\wideparen{B'A'}$. By definition, the phase cohesiveness in Definition
\ref{def:cohesive_c} does not cover this case, since the shortest
arc containing all $(\theta_{A},\theta_{B},\theta_{C})$ and with
length less than $\pi$ does not exist. Let us explain it in terms
of coupling forces. When the angle of bus $C$ is at $C_{2}$, the
coupling forces that applies from $B$ (simply illustrated by $f_{BC}$
in Fig. \ref{fig:example}.a) and that applies from $A$ (illustrated
by $f_{AC}$) counteract with each other. Hence, whether three angles
converge to a common value becomes indeterminate.


Then, consider buses $A$, $B$, $C$ are connected in a way illustrated
in Fig. \ref{fig:example}.c. As known, the coupling forces only exist
between bus $A$ and bus $B$ and between bus $A$ and bus $C$. Suppose
$C$ is at position $C_{2}$ for which phase cohesiveness in Definition
\ref{def:cohesive_c} fails to infer the stability. However, it is
observed that the coupling forces between $A$ and $B$ and between
$A$ and $C$, tends to attract $B$ and $C$ towards $A$, making
the region $\wideparen{BAC_{2}}$ contract and showing it is potentially
stable. \eexample

%
%
%

The observation in Example \ref{exp:angle} motivates us to propose
a different definition of phase cohesiveness that is concerned with
angle differences across lines. Define 
\begin{equation}
\theta_l=B\t\theta, \label{eq:tl}
\end{equation}
where $B$ is the incidence matrices of the  graph $\mathcal{G}$.   Each element in $\theta_{l}$ is angle difference across the corresponding physical line. 
 

\bdefn \label{def:cohesive_l}(Phase Cohesiveness
and Frequency Boundedness). A solution $\theta(t):\mathbb{R}^{+}\rightarrow\mathbb{T}^{n}$
is then said to be phase cohesive if there exists a $\gamma\in[0,\pi)$
such that $\|\theta_l\|_\infty\leq\gamma $. A solution $\dot{
\theta}(t):\mathbb{R}^{+}\rightarrow\mathbb{R}^{n}$
is then said to be frequency bounded if there exists a $\varpi_{o}$
such that $\|\dot{\theta}(t)\|_{\infty}\leq\varpi_{o}$. \edefn

As a result, the second case in Example \ref{exp:angle} could be phase cohesive in the sense of Definition \ref{def:cohesive_l}.
When $\mathcal{G}$ is a complete graph, $\max_{i,j\in\{1,\cdots,n\}}|\theta_{i}-\theta_{j}|\leq\gamma$
is equivalent to $\max_{(i,j)\in\mathcal{V}}|\theta_{i}-\theta_{j}|\leq\gamma$
and therefore Definition \ref{def:cohesive_l}
coincides with Definition \ref{def:cohesive_c}. It
is worth noting that considering the system behavior of $\theta$ in
Euclidean Space $\mathbb{R}^{n}$ and in Torus $\mathbb{T}^{n}$ is
equivalent as far as the initial condition $\theta(t_{o})\in\mathbb{R}^{n}$
at $t=t_{o}$ satisfies $\max_{(i,j)\in\mathcal{E}}|\theta_{i}(t_{o})-\theta_{j}(t_{o})|\leq\gamma$ or $\max_{(i,j)\in\{1,\cdots,n\}}|\theta_{i}(t_{o})-\theta_{j}(t_{o})|\leq\gamma$.

\subsection{Coordinate Transformation and Equilibrium Subspace}
For the operation of  classic power systems and microgrids,  the load demand and the generation of non-dispatchable energy sources are predicted. They are  fed into the optimal power flow algorithm to calculate the  power required to be generated at dispatchable energy sources in order to meet economic goals and system operation requirements. The scheduled power generation matches the predicted demand and their relation is described by  the power flow equation
\begin{equation}
BA_{v}\sin(B\t\theta_e)=p_{o},\label{eq:pf}
\end{equation}
with  
\begin{equation}
\theta_{e}:=\theta_{o}+ce_{n}\label{eq:eq}
\end{equation}
where $p_o$ is a vector consisting of predicted load demand and scheduled power generation satisfying $e_n\t p_o=0$.
$\theta_{o}=\mbox{col}(\theta_{1}^{o},\cdots,\theta_{n}^{o})\in\mathbb{R}^{n}$
is a constant vector  that characterizes the relative angle differences among buses and $c\in\mathbb{R}$ is an arbitrary constant capturing the uniform angle offset on every bus.
Since the uniqueness of equilibria
is fully described by $\theta_{o}$, in what follows, we call
$\theta_{o}$  equilibrium point for simplicity.

In fact, 
 the real-time power profile $p_f$ might not  align 
with the scheduled  $p_{o}$, due to the load and renewable generation fluctuations caused by complicated  load activity and the variation of  the weather condition. Let  $p(t)=p_{f}(t)-p_{o}$
be the power deviation from the dispatched power profile  and regarded as the disturbance to power systems when the system is scheduled to operate around
the equilibrium point $\theta_{o}$. Let 
\begin{equation}
\delta_{i}=\theta_{i}-\theta_{i}^{o}  \label{delta}
\end{equation} be the angle deviation from the equilibrium point. The dynamical system (\ref{eq:FOVO})
can be rewritten in the new coordinate as follows 
\begin{eqnarray}
\dot{\delta} & = & -D^{-1}\left(BA_{v}\left(\sin(B\t(\delta+\theta_{o}))\right.\right.\nonumber \\
 &  & \left.\left.-\sin(B\t\theta_{o})\right)-p(t)\right)\label{eq:FOV}
\end{eqnarray}
where $\delta=[\delta_{1},\cdots,\delta_{n}]\t\in\mathbb{R}^{n}$.
The equilibrium subspace for the system (\ref{eq:FOV})
 is 
 \begin{equation} 
\mbox{\ensuremath{\mathbb{E}}}:=\{\delta\in\mathbb{R}^{n}\mid \delta=ce_{n},\forall c\in\mathbb{R}\} \label{eq:eqsub}
\end{equation}
on which angle deviations are synchronized, i.e., ${\delta}_{i}-{\delta}_{j}=0$,
$\forall i,j=\{1,\cdots,n\}$. As a result, the stability with respect
to an equilibrium point $\theta_{o}$ is converted into the stability
with respect to this equilibrium subspace $\mbox{\ensuremath{\mathbb{E}}}$.

 Denote $\delta_c=B_c\t\delta$   and
\begin{equation}
\bar{\delta}_{c}  = \max_{i,j=1,\cdots,n}\{|\theta_{i}^{o}-\theta_{j}^{o}|\},\label{eq:theta_c} 
\end{equation} 
With the coordinate transformation (\ref{delta}), we present weakened versions of Definition \ref{def:cohesive_c} and \ref{def:cohesive_l}, respectively.

\bdefn \label{def:cohesive_c1} (Synchronization I) A solution $\delta(t):\mathbb{R}^{+}\rightarrow\mathbb{R}^{n}$
is then said to be phase cohesive if there exists a $\gamma\in[0,\pi-\bar{\delta}_{c})$
such that $\|\delta_c(t)\|_\infty\leq\gamma$. A solution
$\dot{\delta}(t):\mathbb{R}^{+}\rightarrow\mathbb{R}^{n}$
is then said to be frequency bounded if there exists a $\varpi_{o}$
such that $\|\dot{\delta}(t)\|_{\infty}<\varpi_{o}$. \edefn

Similarly, denote $\delta_l=B\t \delta$ and
\begin{equation}
\bar{\delta}_{l}  =  \max_{(i,j)\in\mathcal{E}}\{|\theta_{i}^{o}-\theta_{j}^{o}|\}.\label{eq:theta_l}
\end{equation}
The phase cohesiveness and frequency boundedness
in Definition \ref{def:cohesive_l} can
be given in terms of $\delta_l$.
\bdefn \label{def:cohesive_l1} (Synchronization II) A solution $\delta(t):\mathbb{R}^{+}\rightarrow\mathbb{R}^{n}$
is then said to be phase cohesive if there exists a $\gamma\in[0,\pi-\bar{\delta}_{l})$
such that $\|\delta_l(t)\|_\infty\leq\gamma$. A solution
$\dot{\delta}(t):\mathbb{R}^{+}\rightarrow\mathbb{R}^{q}$
is then said to be frequency bounded if there exists a $\varpi_{o}$
such that $\|\dot{\delta}(t)\|_{\infty}<\varpi_{o}$. \edefn

Since $|\theta_{i}^{o}-\theta_{j}^{o}|\leq c_{1}$
and $|\delta_{i}-\delta_{j}|\leq c_{2}$ imply $|\theta_{i}-\theta_{j}|\leq c_{1}+c_{2}$
for $c_{1}+c_{2}\leq\pi$,   the phase cohesiveness and frequency boundedness in Definition \ref {def:cohesive_c1} and Definition \ref {def:cohesive_l1} implies that in  Definition \ref {def:cohesive_c} and \ref {def:cohesive_l}, respectively. Therefore, they are   weakened versions of Definition \ref {def:cohesive_c} and  \ref {def:cohesive_l}.
In fact, taking $(\theta_{o},p_{o})=(0,0)$ in (\ref{eq:pf}) results in $\delta=\theta$, $p=p_{f}$ and $\bar\delta_{l}=\bar\delta_{c}=0$ which in turn
recovers $\theta$-dynamics (\ref{eq:FOVO}) 
from $\delta$-dynamics (\ref{eq:FOV}). In the sequel, we will mainly focus on the 
stability analysis in the sense of of Definition \ref{def:cohesive_c1} and \ref{def:cohesive_l1} and the analysis  can be easily extended to Definition  \ref{def:cohesive_c} and \ref{def:cohesive_l} by taking $(\theta_{o},p_{o})=(0,0)$. In this paper, we have the following assumption.
\bass \label{ass:angle}$\bar\delta_l<\pi/2 $.\eass
Note that this is a reasonable assumption for power systems, since  the secure operation is assured when the angle difference across any physical line is less than $\pi/2$.

\section{Regional Stability Analysis Framework}  \label{sec:SA}
In this section, we will  present a novel regional stability analysis framework that will be applied to   explore the stability of power systems in the sense of Definition \ref{def:cohesive_c1} and \ref{def:cohesive_l1}.
Consider a nonlinear system 
\begin{equation}
\dot{x}=f(t,x)\label{eq:x_system}
\end{equation}
where $x\in\mathbb{R}^{n}$ is the state and the origin
is the equilibrium point of the system (\ref{eq:x_system}), i.e., $f(t,0)=0$.  Define the compact set $B(r):=\{x\in\mathbb{R}^{n}\mid\|x\|\leq r\}$. The analysis is based on the  region-parametrized Lyapunov function defined as follows.

 \bdefn  \label{def:RLF} A continuously differentiable function $V(x):B(r_m)\rightarrow\mathbb{R}^{+}$ is called a region-parametrized Lyapunov function (RPLF) if for any given region $\gamma\in [0,\gamma_{m}]$, there exist  non-negative functions $\underline\alpha$,  $\bar\alpha$ and $\mu$  such that   for $\|x\|\leq\gamma$  it holds that
\begin{gather}
  \underline \alpha(\gamma)\|x\|^2\leq V(x)\leq  \bar\alpha(\gamma) \|x\|^2, \label{eq:V_inq}\\
\frac{\partial V}{\partial x}f(t,x)<0,\;\forall \|x\|\geq   \mu(\gamma). \label{eq:V_dot_Lemma}
\end{gather}
\edefn
\brem
Note that  bounds of the RPLF and the condition for its time derivative to be negative are parametrized by the size $\gamma$ of the region  to be considered.  When the system admits a Lyapunov function  $\tilde V(x): D \rightarrow\mathbb{R}^{+}$ where $ D\in B(r_m)\in \mathbb{R}^n$ satisfying 
$
  \underline \phi ( \|x\|)\leq \tilde V(x)\leq  \bar \phi ( \|x\|)
$
with  class $\mathcal{K}$ functions $\underline \phi$ and $\bar \phi $, we can use $\tilde V$ as the RPLF candidate and explicitly calculate the bounds in (\ref{eq:V_inq}). If $\lim_{s\rightarrow 0}  s^2/\underline \phi(s)< \infty $ and $\lim_{s\rightarrow 0} \bar \phi(s) / s^2 < \infty $,  one can choose
$
\bar\alpha(\gamma)= \sup_{\|x\|\leq\gamma} \{\bar \phi(\|x\|)/\|x\|^2\},\; \underline\alpha(\gamma)= \inf_{\|x\|\leq\gamma} \{\underline \phi(\|x\|)/\|x\|^2\}.
$
\erem
  The next lemma  establishes  the condition on which we can find a positively invariant set within the region $\|x(t)\|\leq \gamma_m$ when there exists a RPLF. It  can be used to investigate the condition for the phase cohesiveness in Definition \ref{def:cohesive_c1} and \ref{def:cohesive_l1}. Before proceeding, let us define the compact set  $W(r):=\{x\in\mathbb{R}^{n}\mid V(x)\leq r\}$ with $V$ as a RPLF.

\blem \label{lem:stab} Consider  nonlinear system (\ref{eq:x_system}). Suppose there exists a RPLF $V(x)$ defined for $\|x\|\leq \gamma_{m}$.  For a given $\gamma\in [0,\gamma_{m}]$, if it holds that
\begin{equation} 
g(\gamma):=\frac{\gamma}{\mu(\gamma)} \sqrt{\frac{\underline\alpha(\gamma)}{ \bar\alpha (\gamma)}} \geq 1,  \label{eq:gamma_inq}
\end{equation} 
 then there exists a $\chi\in\mathbb{R}$ satisfying $ f_l(\gamma) \leq \chi \leq f_r(\gamma)  $
with 
\begin{equation}
f_l(\gamma):= \bar\alpha(\gamma) \mu^2(\gamma),\; f_r(\gamma):= \gamma^2 \underline\alpha(\gamma) \label{eq:flr}
\end{equation}
such that $W(\chi)$ is a positively invariant set, i.e., any trajectories starting with $x(t_{o})\in W(\chi)$
is ultimately contained in $W(f_l(\gamma))$ and  along the trajectory $\|x\|\leq \gamma$ holds.
 \elem
\proofnow See Appendix. \eproof
 
 \bdefn  \label{def:sin} A positive continuous function $g(\gamma): D=\{\gamma|0\leq \gamma\leq\gamma_{m}\} \rightarrow\mathbb{R}^{+}$ is called a quasi-sinusoidal function if
 $ g(0)=0$, $ g(\gamma_{m})=0$
 and $g(\gamma)$ monotonically increases with $x$ for $0\leq \gamma\leq \gamma^*$ and monotonically decreases with $\gamma$  for $\gamma^*<x\leq \gamma_{m}$ where $r^*=\arg\max_{ \gamma\in D}\{g(r)\}$. 
\edefn

The next lemma  gives the sufficient condition under which the solution to the inequality (\ref{eq:gamma_inq}) exists  and further elaborates the result in Lemma \ref{lem:stab} provided that $g(\gamma)$ in (\ref{eq:gamma_inq}) is a quasi-sinusoidal function.

 \blem \label{lem:flr} Consider  nonlinear system (\ref{eq:x_system}) and there exists a RPLF $V(x)$ defined for $\|x\|\leq \gamma_{m}$.   Suppose $g(\gamma)$ in (\ref{eq:gamma_inq}) is a quasi-sinusoidal function of  $\gamma\in[0,\gamma_{m}]$ and maximized at $\gamma=\gamma^*$.   If $g(\gamma^*)>1$, then 
\begin{itemize} 
\item[A.] there exists $0<\gamma_{\min}<\gamma^*$
and $\gamma^*<\gamma_{\max}<\gamma_{m}$ such that $g(\gamma_{\min})=1$,
$g(\gamma_{\max})=1$ and $g(\gamma)>1$
for $\gamma\in(\gamma_{\min},\gamma_{\max})$;
\item[B.]  (Energy Perspective) if $f_l(\gamma)$ is a monotonically increasing function of $\gamma$, $W(\chi)$ is a positively invariant set for every $\chi\in[f_{l,\min},f_{r,\max}]$ where $f_{l,\min}=f_{l} ( \gamma_{\min})$ and $f_{r,\max}=\max_{\gamma\in[\gamma_{\min},\gamma_{\max}]}\{f_{r}(\gamma)\}$. Moreover,
any trajectories starting within $x(t_o) \in W(\chi)$  
is ultimately contained in $W( f_{l,\min})$;
\item[C.] (State Perspective) 
let 
\begin{eqnarray}
\gamma_{l}=\min_{\gamma\in[\gamma_{\min},\gamma_{\max}]}\left\{ \sqrt{\frac{f_{l,\min}}{\underline{\alpha}(\gamma)}}\left|\frac{f_{l,\min}}{\underline{\alpha}(\gamma)}\leq\gamma^{2}\right.\right\},\nonumber\\
\gamma_{r}=\max_{\gamma\in[\gamma_{\min},\gamma_{\max}]}\left\{ \sqrt{\frac{f_{r,\max}}{\bar{\alpha}(\gamma)}}\left|\frac{f_{r,\max}}{\bar{\alpha}(\gamma)}\leq\gamma^{2}\right.\right\}. \label{eq:rls_g}
\end{eqnarray}
Then, any trajectory starting within $B(\gamma)$ 
for $\gamma\in[\gamma_{l},\gamma_{r}]$ is ultimately contained in $B(\gamma_l)$.
\end{itemize} 
\elem

\proofnow See Appendix. \eproof

\section{Synchronization and Transient Stability Analysis}\label{sec:ST}
In this section, we will explore the synchronization of power systems (\ref{eq:FOV}) by proposing a class of parameterized energy functions as the  RPLFs.
\subsection{Energy Functions}
The model of microgrids can be  rewritten 
as 
\begin{equation}
\dot{\delta}=F\delta-G\psi(H\t \delta)+Gp\label{eq:lure}
\end{equation}
where 
\begin{equation}
\psi(H\t x)=BA_{v}\left(\sin(B\t(\delta+\theta_{o}))-\sin(B\t\theta_{o})\right)\label{eq:psi}
\end{equation}
and 
\begin{equation}
F=0, \;G=D^{-1},\;H=I.\label{eq:FGH}
\end{equation}
The equation (\ref{eq:lure}) with $p=0$ is similar to  Lur'e form except that it is not a minimal realization and under-actuated, due to $e_{n}\t\psi(H\t x)=0$.
Let us propose a general class of energy function
\begin{equation}
V(\delta)=V_{1}(\delta)+V_{2}(\delta)\label{eq:V}
\end{equation}
where 
\begin{eqnarray}
V_{1}(\delta) & = &  \frac{1}{2} \alpha \delta\t P \delta \label{eq:V1}\\
V_{2}(\delta) & = & \frac{1}{2}\beta\sum_{i=1}^{n}\sum_{j=1}^{n}a_{ij}\int_{0}^{\delta_{i}-\delta_{j}}\left[\sin(u+\theta_{i}^{o}-\theta_{j}^{o}) \right.\nonumber\\
&&\left.-\sin(\theta_{i}^{o}-\theta_{j}^{o})\right]du.\label{eq:V2}
\end{eqnarray}
with $P\in\mathbb{R}^{n\times n}$ and $\alpha$, $\beta\in\mathbb{R}$
to be determined. 

The following proposition is inspired by the work \cite{Anderson1966,Hill1982} and cited from  \cite{Zhu2016}, which is used to choose $P$, $\alpha$ and $\beta$. 

\bproposition (\cite{Zhu2016}) \label{prop:V} Consider the dynamic system (\ref{eq:lure})
or equivalently the microgrids (\ref{eq:FOV}) 
with $p=0$. If there exist a symmetric matrix $P$ and matrices
$L$, $W$, and $X$ of proper dimensions such that the following
equalities are satisfied
\begin{eqnarray}
PF+F\t P & = & -LL\t\nonumber \\
PG & = & \alpha H+\beta F\t H-LW+Xe_{n}\t\nonumber \\
W\t W & = & \beta(H\t G+G\t H),\label{eq:KYP}
\end{eqnarray}
then the energy function $V$ in (\ref{eq:V}) satisfies $\dot{V}\leq 0$
for $|\delta_{i}-\delta_{j}|<\pi-2\bar{\delta}_{l},\forall(i,j)\in\mathcal{E}$
where $\bar{\delta}_{l}$ is defined in (\ref{eq:theta_l}). \eproposition 
\proofnow The proof is similar to that in  \cite{Anderson1966,Hill1982,Zhu2016} and hence is omitted here. \eproof

According to (\ref{eq:KYP}) and entities in (\ref{eq:FGH}), we follow the procedure presented in \cite{Hill1982} and  find   that $\alpha>0$, $\beta>0$ can be selected arbitrarily and 
\begin{equation}
P=(D-D e_n e_n\t D/d) \label{eq:PP}
\end{equation} 
where $d=e_n\t D e_n$ such that  (\ref{eq:KYP}) is satisfied and the energy function (\ref{eq:V}) obtains $V=0$ at equilibrium subspace
$\mathbb{E}$ defined in (\ref{eq:eqsub}).
It will be shown that the energy function (\ref{eq:V}) is a RPLF for power systems (\ref{eq:FOV}) in the sense of Definition \ref{def:RLF}. Let us define functions
 \begin{equation}
\kappa(\gamma):=\mbox{sinc}(\gamma/2)\cos(\gamma/2+\bar{\delta}_{l}). \label{eq:kappa}
 \end{equation}
 where $\bar{\delta}_{l}$ is given in (\ref{eq:theta_l}) and 
 \begin{equation}
f(\gamma)=\gamma^{q}\kappa^{p}(\gamma)\label{eq:fl}
\end{equation}
Two more lemmas are needed before we proceed to explore the synchronization and transient stability of power systems.
 \blem (\cite{Zhu2016}) \label{lem:fgamma} $\kappa(\gamma)$ is a monotonically decreasing function and  $f(\gamma)$ is a quasi-sinusoidal function in the sense of Definition 
\ref{def:sin}, obtains zeros at $\gamma=0,\pi-2\bar{\delta}_l$
and reaches its maximum at $\gamma=\gamma^*$ 
satisfying 
\begin{equation}
p\cos(\gamma^{*}+\bar{\delta}_l)=(p-q)\kappa(\gamma^{*}).\label{eq:gammao}
\end{equation}
where $\gamma^*< \pi/2-\bar\delta_l$ if $p> q$.
\elem 

\blem \label{lem:lXY}(\cite{Amir-Moez1956}) For Hermitian nonnegative
definite matrices $X$ and $Y$ with ordered eigenvalues, i.e., $\lambda_1(X)\geq\cdots\geq\lambda_{n}(X)$ and $\lambda_1(Y)\geq\cdots\geq\lambda_{n}(Y)$, it holds that
 \begin{gather}
\lambda_{i+j-1}(XY)\leq\lambda_{j}(X)\lambda_{i}(Y),\;i+j\le n+1,\label{eq:lXY_a}\\
\lambda_{i+j-n}(XY)\geq\lambda_{j}(X)\lambda_{i}(Y),\;i+j\geq n+1.\label{eq:lXY_b}
\end{gather}
where $1\leq i,j\leq n$.
\elem
 
\subsection{Synchronization Criterion I} \label{sub:SCI}
In this subsection, we use the  energy function (\ref{eq:V}) with $\alpha=1$, $\beta=0$ and $P$ specified in (\ref{eq:PP}). The energy function is repeated as follows
\begin{equation}
V=\frac{1}{2}\delta\t (D-D e_n e_n\t D/d)\delta \label{eq:Va}.
\end{equation}
Due to $2(D-De_n e_n\t D/d)=\sum_{i=1}^{n}\sum_{j=1}^{n}d_{i}d_{j}(\delta_{i}-\delta_{j})^{2}$, the energy function is similar to  the one used in \cite{Dorfler2012} where the energy function is however defined in the original $\theta$-coordinate. The difference is due to that the synchronization condition  to be derived for power systems (\ref{eq:FOV}) is given in the angle-deviation $\delta$-coordinate  instead of  the original system (\ref{eq:FOVO}). We will adopt the regional stability analysis method presented in Section \ref{sec:SA}. Then,  the synchronization condition in the sense
of Definition \ref{def:cohesive_c1} is presented by the following
theorem with the notation  $B_c(\gamma):=\{\delta_c\in\mathbb{R}^{(n-1)n/2}\mid\|\delta_c\|\leq \gamma\}$.

\bthm \label{thm:robust_c} Consider power systems (\ref{eq:FOV}) with energy function (\ref{eq:Va}) under Assumption \ref{ass:angle}. Assume $\bar\delta_c-\bar\delta_l<\pi/2$. Let $\gamma\in[0,\pi-\bar\delta_{m})$ where $\bar\delta_m=\max\{2\bar{\delta}_{l},\bar{\delta}_c\}$ and $\kappa(\gamma)$ be defined in (\ref{eq:kappa}). Let function
$\sigma(\gamma)$\footnote{For convenience, we denote $\|s(t)\|_{[t_1,t_2]}=\sup_{t_1\leq t\leq t_2}\|s(t)\|$ for a bounded vector singal $s(t)$. } be
\begin{eqnarray}
\sigma(\gamma)&=&\sqrt{\max_{i\neq j}\{d_i d_j\}/\min_{i\neq j}\{d_i d_j\}}\nonumber\\
&&\times\frac{n \|\mbox{diag}\{d_i d_j\} B_c\t D^{-1}p(t)\|_{[t_0,\infty]}}{\gamma \kappa(\gamma) d}.
\end{eqnarray}
If it holds that 
\begin{equation}
 \lambda_2>\lambda_{cr}:=\sigma(\gamma^*),\label{eq:sigma_varep}
\end{equation}
where $\lambda_2$ is the algebraic connectivity of the underlying Laplacian $L$ of power network and $\gamma^{*}=\pi/2-\bar\delta_l$, the synchronization in the sense of Definition \ref{def:cohesive_c1} is achieved. In particular, 
\begin{itemize} 
\item[A.] there exists $0<\gamma_{\min}<\pi/2-\bar\delta_l$
and $\pi/2-\bar\delta_l<\gamma_{\max}<\pi-\bar{\delta}_{m}$ such that $ \lambda_2=\sigma(\gamma_{\min})$,
$ \lambda_2=\sigma(\gamma_{\max})$ and $\lambda_2>\sigma(\gamma)$
for $\gamma\in(\gamma_{\min},\gamma_{\max})$;
\item[B.] (Energy Perspective) $W(\chi)$ is a positively invariant set for any $\chi\in[ f_{l,\min},f_{r,\max}]$ where 
\begin{eqnarray}
f_{l,\min}=  \gamma^2_{\min} \min_{i\neq j}\{d_i d_j\}/(2d),\label{flmin_c} \\
 f_{r,\max}=\gamma_{\max}^2 \min_{i\neq j}\{d_i d_j\}/(2d), \label{frmax_c}
 \end{eqnarray} i.e.,
any trajectories $\delta$ starting within $x(t_o) \in W(\chi)$  
is ultimately contained in $W( f_{l,{\min}})$;
\item[C.] (Angle Perspective: phase cohesiveness) 
let   
\begin{equation}
\gamma_l=\gamma_{\min},
\gamma_r=\gamma_{\max}\sqrt{\frac{\min_{i\neq j}\{d_i d_j\}}{\max_{i\neq j}\{d_i d_j\}}}.
\end{equation}
Then, any trajectories $\delta(t)$ starting within $B_c(\gamma)$ for $\gamma\in[\gamma_l,\gamma_r]$ is ultimately contained in $B_c(\gamma_l)$.
\item[D.] (Frequency Boundedness) if $ p$ is bounded, there exists a $T$ such that $\|\dot \delta \|_{\infty}\leq\varpi_{o}$ for
some $\varpi_{o}$ and $t>T$.
\end{itemize} 
 \ethm 
\proofnow
First, we will verify the energy function $V(\delta)$ in (\ref{eq:Va}) is a RPLF.  Note that 
\begin{equation}
(D-De_n e_n\t D/d)= B_c \mbox{diag} \{d_i d_j\} B_c\t /d  \label{eq:DBc}
\end{equation}
where $B_{c}$ is the incidence matrix of the induced complete graph. It leads to 
\begin{equation}
\underline\alpha \|\delta_c\| \leq V\leq\bar\alpha \|\delta_c\|
\end{equation}
where $\delta_c=B_c \delta$ and 
\begin{equation}
\underline\alpha= \min_{i\neq j}\{d_i d_j\}/(2d),\;\bar\alpha= \max_{i\neq j}\{d_i d_j\}/(2d). \label{eq:alph}
\end{equation}
For the rest of the proof, we consider Lyapunov function $V$ within $\|\delta_c\|\leq \gamma$ for $\gamma\in[0,\pi-2\bar\delta_{l})$. Note that $\|\delta_l\|_\infty \leq \|\delta_c\|_\infty \leq \|\delta_c\|\leq \gamma$ implies that $|\delta_i-\delta_j|\leq \gamma$, $\forall (i,j)\in\mathcal{E}$.
The derivative of $V(\delta)$, along the trajectory of
(\ref{eq:FOV}), is 
\begin{eqnarray}
\dot{V}&=& \delta\t(I-De_{n}e_{n}\t/d)[p  \nonumber\\
&&-BA_{v}(\sin(B\t(\delta+\theta_{o}))-\sin(B\t\theta_{o}))]\nonumber\\
&=&-\delta\t BA_{v}(\sin(B\t(\delta+\theta_{o}))-\sin(B\t\theta_{o})) \nonumber\\
&&+\delta\t(I-De_{n}e_{n}\t/d)p.\label{eq:V_temp}
\end{eqnarray}
Since $|\delta_{i}-\delta_{j}|\leq\gamma$ and $\|\theta_{i}^{o}-\theta_{j}^{o}\|\leq\bar\delta_{l}$ for any bus $j$ that is connected with bus $i$,
one has 
\begin{align*}
\frac{\sin(\delta_{i}-\delta_{j}+\theta_{i}^{o}-\theta_{j}^{o})-\sin(\theta_{i}^{o}-\theta_{j}^{o})}{\delta_{i}-\delta_{j}}\\
=\frac{\cos((\delta_{i}-\delta_{j})/2+\theta_{i}^{o}-\theta_{j}^{o})\sin((\delta_{i}-\delta_{j})/2)}{(\delta_{i}-\delta_{j})/2}\geq\kappa(\gamma) \label{eq:kappa_def}
\end{align*}
where $\kappa(\gamma)$ is defined in (\ref{eq:kappa}).
As a result, 
\begin{gather*}
\delta\t BA_{v}(\sin(B\t(\delta+\theta_{o}))=\frac{1}{2}\sum_{i=1}^{n}\sum_{j=1}^{n}\left[a_{ij}(\delta_{i}-\delta_{j})^{2}\right.\\
\left.\frac{\sin(\delta_{i}-\delta_{j}+\theta_{i}^{o}-\theta_{j}^{o})-\sin(\theta_{i}^{o}-\theta_{j}^{o})}{\delta_{i}-\delta_{j}}\right]\\
\geq\kappa(\gamma)\delta\t\mathcal{L}\delta\geq\frac{\lambda_{2}}{n}\kappa(\gamma)\|\delta_{c}\|^{2},
\end{gather*}
where the last inequality is due to Lemma 4.7 in \cite{Dorfler2012} and 
\[
\delta\t(I-De_{n}e_{n}\t/d)p=\delta_{c}\mbox{diag}\{d_{i}d_{j}\}B_{c}\t D^{-1}p/d.
\]
where we used (\ref{eq:DBc}). 
Equation (\ref{eq:V_temp}) leads to
\begin{align*}
\dot{V}&\leq-\frac{\lambda_{2}}{n}\kappa(\gamma)\|\delta_{c}\|^{2}+\delta_{c}\mbox{diag}\{d_{i}d_{j}\}B_{c}\t D^{-1}p/d\\
&\leq0,\;\mbox{if}\;\|\delta_{c}\|\geq\frac{n\|\mbox{diag}\{d_{i}d_{j}\}B_{c}\t D^{-1}p(t)\|_{[t_0,\infty]}}{\kappa(\gamma) \lambda_{2}d}.
\end{align*}
So far, we concluded that $V$ is a RPLF.
Then, 
if 
\begin{equation}
 g(\gamma):=\gamma \kappa(\gamma)/R_s\geq 1, \label{eq:gamma_ine}
\end{equation}
where 
\begin{align}
R_s:=&\sqrt{\max_{i\neq j}\{d_i d_j\}/\min_{i\neq j}\{d_i d_j\}}\nonumber\\
&\times\frac{n \|\mbox{diag}\{d_i d_j\} B_c\t D^{-1}p(t)\|_{[t_0,\infty]}}{\lambda_2 d},
\end{align}
the condition of Lemma \ref{lem:stab} is satisfied. We further analyze the inequality (\ref{eq:gamma_ine}) using Lemma \ref{lem:flr}. Note that $g(\gamma)$ in (\ref{eq:gamma_ine}) is a quasi-sinusoidal function and maximizes at $\gamma^*=\pi/2-\bar{\delta}_l$ by Lemma \ref{lem:fgamma}. If $g(\gamma^*)>1$ which is equivalent to (\ref{eq:sigma_varep}), it follows from Statement A of Lemma \ref{lem:flr} that Statement A is satisfied. We can calculate  $f_{l,\min}$ in Lemma \ref{lem:flr} as
\begin{equation}
f_{l,\min}=  \frac{n^2\|\mbox{diag}\{d_{i}d_{j}\}B_{c}\t D^{-1}p(t)\|_{[t_0,\infty]}^2  \max_{i\neq j}\{d_i d_j\}}{2\kappa^2(\gamma_{\min}) \lambda_{2}^2 d^3} 
\end{equation}
and $f_{r,\max}=\underline \alpha \gamma_{\max}^2$.
Noting  $ \lambda_2=\sigma(\gamma_{\min})$ and $\underline \alpha$ in (\ref{eq:alph}), we can obtain the neat expression of $f_{l,\min}$ in (\ref{flmin_c}) and $f_{r,\max}$ in (\ref{frmax_c}). As a result, Statement B follows  that of Lemma \ref{lem:flr}. Statement C follows from that of Lemma \ref{lem:flr} by noting functions  $\underline\alpha(\gamma)$ and $\bar\alpha(\gamma)$ do not depend on $\gamma$.

What remains is to prove  frequency boundedness. 
Statement C implies that there exists a 
$T$ such that  the system trajectory $\delta_c(t)$ enters and stay in the ball $B(\gamma_l)$ for $t>T$ where $\gamma_l=\gamma_{\min} < \gamma^* =\pi/2-\bar\delta_l$. We also note $\|\delta_c\|\leq\gamma_l=\gamma_{\min}<\pi/2-\bar\delta_l$  implies that $|\delta_{i}-\delta_{j}+\theta_{ij}^{o}|<\frac{1}{2}\pi$.  As a result, RHS of (\ref{eq:FOV}) is bounded, which shows that $\|\dot \delta\|$ is bounded.
So, the frequency
$\dot{\delta}$ will be ultimately bounded, i.e., $\|\dot{\delta}(t)\|\leq\varpi_{o}$ for $t>T$
with some $\varpi_{o}$ and $T$. The frequency boundedness
is proved. 
\eproof

Taking $(\theta_{o},p_{o})=(0,0)$ in (\ref{eq:pf})
recovers $\theta$-dynamics (\ref{eq:FOVO}) 
from $\delta$-dynamics (\ref{eq:FOV}) and makes $\kappa(\gamma)=\sin(\gamma)/\gamma$. As a result, the energy function becomes $V=\theta\t(D-D e_n e_n\t D/d)\theta$ which coincides with  the one used in \cite{Dorfler2012}. Then, we arrive at the following corollary with this energy function.
\bcorollary \label{cor:robust} Consider microgrid (\ref{eq:FOVO}) with
 energy function $V=\theta\t(D-D e_n e_n\t D/d)\theta$. Let $\gamma\in[0,\pi)$
 and function $\bar\sigma(\gamma)$ be
\begin{eqnarray*}
\bar\sigma(\gamma)&=&\sqrt{\max_{i\neq j}\{d_i d_j\}/\min_{i\neq j}\{d_i d_j\}}\nonumber \\
&&\times\frac{n \|\mbox{diag}\{d_i d_j\} B_c\t D^{-1}p(t)\|_{[t_0,\infty]}}{\sin(\gamma) d}
\end{eqnarray*}
If $ \lambda_2>\lambda_{cr}:=\bar \sigma(\gamma^*)$ holds
where $\gamma^{*}=\pi/2$, the synchronization in the sense of Definition \ref{def:cohesive_c} is achieved. In particular, 
\begin{itemize} 
\item[A.] there exists $0<\gamma_{\min}\leq\pi/2$
and $\pi/2<\gamma_{\max}<\gamma_{\max}$ such that $ \lambda_2=\bar\sigma(\gamma_{\min})$,
$ \lambda_2=\bar\sigma(\gamma_{\max})$ and $\lambda_2>\lambda_{cr}$
for $\gamma\in(\gamma_{\min},\gamma_{\max})$;
\item[B.] (Energy Perspective) $W(\chi)$ is a positively invariant set for any $\chi\in[ f_{l,\min},f_{r,\max}]$ where  
\begin{eqnarray}
f_{l,\min}=  \gamma^2_{\min} \min_{i\neq j}\{d_i d_j\}/(2d),\label{eq:flmin} \\
 f_{r,\max}=\gamma_{\max}^2 \min_{i\neq j}\{d_i d_j\}/(2d), \label{eq:frmax}
 \end{eqnarray} i.e.,
any trajectories $\theta$ starting within $x(t_o) \in W(\chi)$  
is ultimately contained in $W( f_{l,{\min}})$;
\item[C.] (Angle Perspective: phase cohesiveness) 
let   
\begin{equation}
\gamma_l=\gamma_{\min},
\gamma_r=\gamma_{\max}\sqrt{\frac{\min_{i\neq j}\{d_i d_j\}}{\max_{i\neq j}\{d_i d_j\}}}.
\end{equation}
Any trajectories $\theta$ starting within $B_c(\gamma)$ for  $\gamma\in[\gamma_l,\gamma_r]$ is ultimately contained in $B_c(\gamma_l)$.
\item[D.] if $ p$ is bounded, there exists a $T$ such that $\|\dot \theta \|_{\infty}\leq\varpi_{o}$ for
some $\varpi_{o}$ and $t>T$.
\end{itemize} 
 \ecorollary 
 \brem
Statement A and C in Corollary \ref{cor:robust}  coincide with Theorem 4.4 in \cite{Dorfler2012} where  constant power profile $p_f$ is considered and frequency synchronization can be achieved. As we consider some entries in power profile are time-varying, Corollary \ref{cor:robust} extends  the  result in \cite{Dorfler2012} to frequency boundedness in Statement D  and  in addition provide the existence of the invariant set  from energy perspective in Statement B. Also,  the condition is derived using  the regional stability analysis method proposed in Section \ref{sec:SA}.
\erem

\subsection{Synchronization Criterion II} \label{sub:SCII}
In this subsection, we use the  energy function (\ref{eq:V}) with $\alpha=0$ and $\beta=1$ which is repeated as follows
\begin{equation}
V(\delta)=\frac{1}{2}\sum_{i=1}^{n}\sum_{j=1}^{n}a_{ij}\int_{0}^{\delta_{i}-\delta_{j}}(\sin(u+\theta_{ij}^{o})-\sin\theta_{ij}^{o})du.\label{eq:Vl}
\end{equation}
where $\theta_{ij}^{o}=\theta_{i}^{o}-\theta_{j}^{o}$. In fact, $V(\delta)$
is the sum of the potential energy induced by the coupling force between
$i$th bus and $j$th bus when angles  move away from the equilibrium $\theta_{o}$.
Since $a_{ij}\neq0$ if and only if $(i,j)\in\mathcal{E}$,
$V(\delta)$ sums up the potential energy only induced across  transmission lines.

\blem \label{lem:V} For a given $\gamma\in[0,\pi-2\bar\delta_{l})$, if $\max_{(i,j)\in\mathcal{E}}|\delta_{i}-\delta_{j}|\leq\gamma$,
then 
\begin{gather}
\underline{\alpha}(\gamma)\|\delta_l\|^{2}\leq V(\delta)\leq\bar{\alpha}\|\delta_l\|^{2}\label{eq:V2_bound-1}
\end{gather}
holds for
\begin{equation}
\underline{\alpha}(\gamma)=\kappa(\gamma)\min_{(i,j)\in\mathcal{E}}\{a_{ij}\}/2,\;
\bar{\alpha}=\max_{(i,j)\in\mathcal{E}}\{a_{ij}\}/2.
\end{equation}
\elem
\proofnow 
For any $|u|\leq\gamma<\pi$
and $|x|\leq\delta_{l}$, one has 
\begin{equation}
\frac{\sin(u+x)-\sin x}{u}\leq1.\label{eq:u_delta1}
\end{equation}
Applying (\ref{eq:kappa_def}) and (\ref{eq:u_delta1})  yields
\begin{gather*}
V(\delta)=\frac{1}{2}\sum_{i=1}^{n}\sum_{i=1}^{n}a_{ij}\int_{0}^{\delta_{i}-\delta_{j}}\frac{\sin(u+\theta_{ij}^{o})-\sin\theta_{ij}^{o}}{u}udu\\
\geq\frac{1}{2}\sum_{i=1}^{n}\sum_{i=1}^{n}a_{ij}\int_{0}^{\delta_{i}-\delta_{j}}\kappa(\gamma)udu=\frac{1}{2}\kappa(\gamma)\delta\t BA_{v}B\t\delta
\end{gather*}
and 
\[
V(\delta)\leq\frac{1}{2}\delta\t BA_{v}B\t\delta
\]
Noting $\delta_l=B\t\delta$, the proof is complete. 
\eproof

Lemma \ref{lem:V} shows that the energy function $V(\delta)$ is
bounded by quadratic functions of $\|\delta_l\|$. Using $V(\delta)$
as the RPLF candidate, the synchronization condition in the sense
of Definition \ref{def:cohesive_l1} is presented by the following
theorem with  $B_l(\gamma):=\{\delta_l\in\mathbb{R}^{|\mathcal{E}|}\mid\|\delta_l\|\leq \gamma\}$.

\bthm \label{thm:robust_l}  Consider power systems (\ref{eq:FOV}) with
 energy function (\ref{eq:Vl}) under Assumption \ref{ass:angle}. Let $\gamma\in[0,\pi-2\bar\delta_{l})$
  and $\kappa(\gamma)$ be defined in
(\ref{eq:kappa}). Let function $\sigma(\gamma)$ be 
\begin{equation}
\sigma(\gamma):=\frac{\|A_{v}B\t D^{-1}p(t)\|_{[t_0,\infty]}}{\gamma\kappa^{\frac{5}{2}}(\gamma)}\left(\frac{\max_{(i,j)\in\mathcal{E}}\{a_{ij}\}}{\min_{(i,j)\in\mathcal{E}}\{a_{ij}\}}\right)^{\frac{3}{2}}.
\end{equation}
Define 
\begin{equation}
Q=A_{v}B\t D^{-1}BA_{v}\geq0.\label{eq:Q}
\end{equation}
If it holds that 
\begin{equation}
\lambda_{s}(Q)>\lambda_{cr}:=\sigma(\gamma^*), \label{eq:sigma_varepl}
\end{equation}
where $\lambda_{s}(Q)$ is the smallest non-zero eigenvalue of $Q$ and $\gamma^*$ satisfies
\begin{equation}
\kappa(\gamma^{*})=\frac{5}{3}\cos(\gamma^{*}+\bar{\delta}_l), \label{eq:gamma_starl}
\end{equation}
then the synchronization in the sense of Definition \ref{def:cohesive_l1} is achieved. In particular, 
\begin{itemize} 
\item[A.] there exists $0<\gamma_{\min}<\gamma^*$
and $\gamma^*<\gamma_{\max}<\gamma_{\max}$ such that $ \lambda_s(Q)=\sigma(\gamma_{\min})$,
$ \lambda_s(Q)=\sigma(\gamma_{\max})$ and $\lambda_s(Q)>\lambda_{cr}$
for $\gamma\in(\gamma_{\min},\gamma_{\max})$;
\item[B.] (Energy Perspective) $W(\chi)$ is a positively invariant set for any $\chi\in[ f_{l,\min},f_{r,\max}]$ where 
\begin{eqnarray}
f_{l,\min}= \gamma^2_{\min}\kappa(\gamma_{\min})\min_{(i,j)\in\mathcal{E}}\{a_{ij}\}/2, \label{flmin_l} \\
 f_{r,\max}=\gamma^2_{s}\kappa(\gamma_{s})\min_{(i,j)\in\mathcal{E}}\{a_{ij}\}/2, \label{frmax_l}
 \end{eqnarray} 
where  $\gamma_s$ satisfies
\begin{equation}
\cos(\gamma_s+\bar\delta_l)=-\kappa(\gamma_s), \label{eq:gamma_sl}
\end{equation}
i.e., any trajectories $\delta$ starting within $x(t_o) \in W(\chi)$ where 
is ultimately contained in $W( f_{l,{\min}})$;
\item[C.] (Angle Perspective: phase cohesiveness) 
let   
\begin{equation}
\gamma_l=\gamma_{\min},
\gamma_r=\gamma_{s}\sqrt{\kappa(\gamma_s)\frac{\min_{(i,j)\in\mathcal{E}}\{a_{ij}\}}{\max_{(i,j)\in\mathcal{E}}\{a_{ij}\}}}.
\end{equation}
Then, any trajectories $\delta$ starting within $B_l(\gamma)$ for $\gamma\in[\gamma_l,\gamma_r]$ is ultimately contained in $B_l(\gamma_l)$.
\item[D.]  (Frequency Boundedness) if $ p$ is bounded,  there exists a $T$ such that $\|\dot \delta \|_{\infty}\leq\varpi_{o}$ for
some $\varpi_{o}$ and $t>T$.
\end{itemize} 
 \ethm 

\proofnow Let us consider Lyapunov function  (\ref{eq:Vl}) within $\|\delta_l\|\leq \gamma$ for $\gamma\in[0,\pi-2\bar\delta_{l})$. The derivative of $V(\delta)$, along the trajectory of
(\ref{eq:FOV}), is 
\begin{gather}
\dot{V}(\delta)=\frac{1}{2}\sum_{i=1}^{n}\sum_{j=1}^{n}a_{ij}(\sin(\delta_{i}-\delta_{j}+\theta_{ij}^{o})-\sin\theta_{ij}^{o})(\dot{\delta}_{i}-\dot{\delta}_{j})\nonumber\\
=S(\delta)+Q(\delta)  \label{eq:SQ}
\end{gather}
 where $S(\delta)$ and $Q(\delta)$ are denoted as 
\begin{gather*}
S(\delta)=\frac{1}{2}\sum_{i=1}^{n}\sum_{j=1}^{n}a_{ij}(\sin(\delta_{i}-\delta_{j}+\theta_{ij}^{o})-\sin\theta_{ij}^{o})\\
\times\left\{ -d_{i}^{-1}\sum_{k=1}^{n}a_{ik}\left(\sin(\delta_{i}-\delta_{k}+\theta_{ik}^{o})-\sin(\theta_{ik}^{o})\right)\right.\\
\left.+d_{j}^{-1}\sum_{k=1}^{n}a_{jk}\left(\sin(\delta_{j}-\delta_{k}+\theta_{jk}^{o})-\sin(\theta_{jk}^{o})\right)\right\} 
\end{gather*}
and 
\[
Q(\delta)=\frac{1}{2}\sum_{i=1}^{n}\sum_{j=1}^{n}a_{ij}(\sin(\delta_{i}-\delta_{j}+\theta_{ij}^{o})-\sin\theta_{ij}^{o})(d_{i}^{-1}p_{i}-d_{j}^{-1}p_{j})
\]
A manipulation of the indices in $S(\delta)$  leads to
\begin{gather}
S(\delta)
=-\sum_{i=1}^{n}\frac{1}{d_{i}}\left\{ \sum_{j=1}^{n}a_{ij}(\sin(\delta_{i}-\delta_{j}+\theta_{ij}^{o})-\sin\theta_{ij}^{o})\right\} \times \nonumber \\
\left\{ \sum_{j=1}^{n}a_{ij}\left(\sin(\delta_{i}-\delta_{j}+\theta_{ij}^{o})-\sin(\theta_{ij}^{o})\right)\right\} \label{eq:S_delta}
\end{gather}

The RHS of the last equality in (\ref{eq:S_delta})
equals to 
\[
S(\delta)=-\delta\t BA_{p}A_{v}B\t D^{-1}BA_{v}A_{p}B\t\delta
\]
where $A_{p}$ is a diagonal matrix with diagonal elements being 
\[
A_{p}(k,k)=\frac{\sin(\delta_{i}-\delta_{j}+\theta_{ij}^{o})-\sin\theta_{ij}^{o}}{\delta_{i}-\delta_{j}},
\]
for  $(i,j)=\mathcal{E}_k, k=1,\cdots,|\mathcal{E}|$.
It is noted from (\ref{eq:kappa_def}) that $A_{p}(k,k)\geq\kappa(\gamma)$,
when $|\theta_{i}^{o}-\theta_{j}^{o}|\leq\delta_{l}$ and $|\delta_{i}(t)-\delta_{j}(t)|<\gamma$
for any $(i,j)\in\mathcal{E}$. 
One
has 
\begin{gather}
\delta\t BA_{p}A_{v}B\t D^{-1}BA_{v}A_{p}B\t\delta=
\xi\t\bar{Q} \xi  \nonumber \\
\geq\|\xi\|^{2}\min_{\xi \neq 0}\frac{\xi \t \bar{Q} \xi}{\xi\t\xi}. \label{eq:mid_cal}
\end{gather}
where $\xi=(A_{v}A_{p})^{\frac{1}{2}} B\t\delta$ and $\bar Q=(A_{p}A_{v})^{\frac{1}{2}} B\t D^{-1}B (A_{v} A_{p})^{\frac{1}{2}}$.  Note that $\bar Q$ is the symmetric matrix whose eigenvalues are all non-negative. Let $v$ be the eigenvector corresponding to the zero eigenvalue of $\bar Q$. We note that $v\in \mbox{null}(\bar Q)\in \mbox{null}(B(A_v A_p)^{\frac{1}{2}})$, since $\bar Q v=0$ implies $ B(A_v A_p)^{\frac{1}{2}}v=0$. Also, $\xi \in \mbox{imag}((A_{v}A_{p})^{\frac{1}{2}} B\t)$ which shows that $\xi \perp v$, because the null space of $B(A_v A_p)^{\frac{1}{2}}$ is orthogonal  complement to the column space of  $(A_{v}A_{p})^{\frac{1}{2}} B\t$. 
Then, by Courant-Fischer minimum-maximum theorem, one has 
\[
\min_{\xi \neq 0}\frac{\xi \t \bar{Q} \xi}{\xi\t\xi}=\min_{\xi \neq 0, \xi \perp v}\frac{\xi \t \bar{Q} \xi}{\xi\t\xi}=\lambda_s(\bar Q)
\]
where $\lambda_{s}(\bar Q)$ is the smallest non-zero eigenvalue of $\bar Q$. Also, $\|\xi\|^2 \geq \kappa(\gamma)\min_{(i,j)\in\mathcal{E}}\{a_{ij}\}\|\delta_l\|^2$ by noting $\delta_l=B\t \delta$. Due to $\bar Q=(A^{-1}_{v} A_{p})^{\frac{1}{2}}  Q (A^{-1}_{v} A_{p})^{\frac{1}{2}}$,
$\lambda_s (\bar Q)\geq \lambda_s(Q) \kappa(\gamma)/\max_{(i,j)\in\mathcal{E}}\{a_{ij}\}$ by Lemma \ref{lem:lXY}.
From (\ref{eq:mid_cal}),
\begin{align*}
\delta\t BA_{p}A_{v}B\t D^{-1}BA_{v}A_{p}B\t\delta\geq \\
\frac{\min_{(i,j)\in\mathcal{E}}\{a_{ij}\}}{\max_{(i,j)\in\mathcal{E}}\{a_{ij}\}}  \lambda_s(Q) \kappa^2(\gamma) \|\delta_l\|^2
\end{align*}
 

Also one has  $Q(\delta)$ in (\ref{eq:SQ})
\begin{gather*}
Q(\delta)=\frac{1}{2}\sum_{i=1}^{n}\sum_{j=1}^{n}a_{ij}\frac{(\sin(\delta_{i}-\delta_{j}+\theta_{ij}^{o})-\sin\theta_{ij}^{o})}{\delta_{i}-\delta_{j}}(\delta_{i}-\delta_{j})\\
\times(d_{i}^{-1}p_{i}-d_{j}^{-1}p_{j})=\delta\t BA_{v}A_{p}B\t D^{-1}p\\
\leq\|A_v B\t D^{-1}p(t)\|_{[t_0,\infty]}\|\delta_l\|
\end{gather*}
due to $\|A_{p}\|<1$, Thus, $\dot{V}$ is bounded by 
\begin{align}
\dot{V}\leq &- \kappa^2(\gamma) \lambda_s(Q) \frac{\min_{(i,j)\in\mathcal{E}}\{a_{ij}\}}{\max_{(i,j)\in\mathcal{E}}\{a_{ij}\}} \|\delta_l\|^{2}\nonumber\\
&+\|A_v B\t D^{-1}p(t)\|_{[t_0,\infty]}\|\delta_l\| \nonumber\\
\leq &0,\;\mbox{if}\; \|\delta_l\|\geq  \frac{\|A_v B\t D^{-1}p(t)\|_{[t_0,\infty]} \max_{(i,j)\in\mathcal{E}}\{a_{ij}\}}{\lambda_s(Q) \kappa^2(\gamma) \min_{(i,j)\in\mathcal{E}}\{a_{ij}\}}   \label{eq:V_dot_ineq}
\end{align}
So far, we proved that $V$ is a RPLF. Then, 
if 
\begin{equation}
g(\gamma):=\gamma \kappa^{\frac{5}{2}}(\gamma)/R_s>1, \label{eq:gamma_inel}
\end{equation}
where 
\begin{equation}
R_s:=\left(\frac{\max_{(i,j)\in\mathcal{E}}\{a_{ij}\}}{\min_{(i,j)\in\mathcal{E}}\{a_{ij}\}}\right)^{\frac{3}{2}} \frac{ \|A_v B\t D^{-1}p(t)\|_{[t_0,\infty]}}{\lambda_s(Q)}
\end{equation}
the condition of Lemma \ref{lem:stab} is satisfied. We further analyze the inequality (\ref{eq:gamma_inel}) using Lemma \ref{lem:flr}. Note that $g(\gamma)$ in (\ref{eq:gamma_ine}) is a quasi-sinusoidal function by Lemma \ref{lem:fgamma} and maximizes at $\gamma^*$ which satisfies (\ref{eq:gamma_starl}). If $g(\gamma^*)>1$ which is equivalent to (\ref{eq:sigma_varepl}), it follows from Lemma \ref{lem:flr} that Statement A is satisfied with $\gamma_{\min}< \pi/2-\bar\delta_l$ (by Lemma \ref{lem:fgamma}). We calculate 
  $f_l(\gamma)$ as follows 
\[
f_{l}(\gamma)=\frac{\|A_{v}B\t D^{-1}p(t)\|_{[t_0,\infty]}^2\left(\max_{(i,j)\in\mathcal{E}}\{a_{ij}\}\right)^{3}}{2\lambda_{s}^{2}(Q)\kappa^{4}(\gamma)\left(\min_{(i,j)\in\mathcal{E}}\{a_{ij}\}\right)^{2}}
\]
It is a monotonically increasing function of $\gamma$, because $\kappa(\gamma)$ is a monotonically decreasing function by Lemma \ref{lem:fgamma}. As a result,
$
f_{l,\min}= f_{l}(\gamma_{\min}) $.  We can obtain the neat expression of $f_{l,\min}$ in (\ref{flmin_l}) by noting $ \lambda_s(Q)=\sigma(\gamma_{\min})$.  
We can calculate $f_r(\gamma)=\gamma^2 \kappa(\gamma)\min_{(i,j)\in\mathcal{E}}\{a_{ij}\}/2$. By Lemma \ref{lem:fgamma},  
\[
\arg\max_{\gamma\in[\gamma_{\min},\gamma_{\max}]}\{\gamma^{2}\kappa(\gamma)\}=\gamma_{s},
\]
where $\gamma_s$ satisfies (\ref{eq:gamma_sl}).
Thus, Statement B follows. 
Applying Lemma \ref{lem:flr} shows  $\gamma_{l}$  is
$\gamma_{l}=\gamma_{\min}.$
Then, Statement C follows from that of Lemma \ref{lem:flr} by noting that the function $\bar\alpha$ does not depend on $\gamma$. 
Statement D about  frequency boundedness can easily follows from similar argument in Theorem \ref{thm:robust_c} by noting that the system trajectory is ultimately contained in $B(\gamma_l)$, i.e., $\|\delta_l\|\leq\gamma_{l}=\gamma_{\min}\leq\pi/2-\bar\delta_l$.
\eproof

\section{Extension to Non-disturbance Case: Region of Attraction} \label{sec:ext}
In this section, we will extend the regional stability analysis  method presented in Section \ref{sec:SA} and stability analysis in Section \ref{sub:SCI} and \ref{sub:SCII}   to the non-disturbance case, i.e., $p=0$ in (\ref{eq:FOV}). These results can be utilized to estimate the region of attraction for power systems (\ref{eq:FOV}), which is useful to assess the stability of power systems following severe faults such as tripping of a line. We first derive a variant of  Lemma \ref{lem:flr} when the derivative of the RPLF $V(x)$ satisfies (\ref{eq:V_dot_Lemma}) with $\mu(\gamma)=0$.
 \blem \label{lem:flr_e} Consider  nonlinear system (\ref{eq:x_system}) with regional Lyapunov function  satisfying (\ref{eq:V_inq}) and  (\ref{eq:V_dot_Lemma}) with $\mu(\gamma)=0$. Then,
 \begin{itemize} 
\item[A.]  (Energy Perspective) $W(\chi)$ is a positively invariant set for any $\chi\in[0,f_{r,\max}]$ where  \begin{equation}
f_{r,\max}=\max_{\gamma\in[0,\gamma_{m}]}\{ \gamma^2 \underline\alpha(\gamma)\},
\end{equation}
i.e., any trajectories starting within $x(t_o) \in W(\chi)$  ultimately converge to equilibrium point.
\item[B.] (State Perspective)  
let    
\begin{equation}
\gamma_{r}=\max_{\gamma\in[0,\gamma_{m}]}\left\{ \sqrt{\frac{f_{r,\max}}{\bar{\alpha}(\gamma)}}\left|\frac{f_{r,\max}}{\bar{\alpha}(\gamma)}\leq\gamma^{2}\right.\right\}.\label{eq:rls_l}
\end{equation}
Then, any trajectories starting within $B(\gamma_r)$ for $\gamma\in[0,\gamma_r]$ ultimately converge to equilibrium   point, i.e., the region of attraction is $\|x\|\leq \gamma_r$.
\end{itemize} 
\elem
\proofnow Note that $f_r(\gamma)>f_l(\gamma)$ in  Lemma \ref{lem:stab} holds for all $\gamma\in[0,\gamma_{m}]$, since $\mu(\gamma)=0$. The proof easily follows from that of Lemma \ref{lem:stab} and \ref{lem:flr} by setting $\mu(\gamma)=0$. \eproof

Then, we can use Lemma \ref{lem:flr_e} to show the next two theorems. 
\bthm \label{thm:robust_ce} Consider microgrid (\ref{eq:FOV}) with $p(t)=0$ and
 energy function (\ref{eq:Va}).   
\begin{itemize} 
\item[A.] (Energy Perspective) $W(\chi)$ is a positively invariant set for any $\chi\in[ 0,f_{r,\max}]$ where 
\begin{equation}
 f_{r,\max}=(\pi-\bar\delta_m)^2 \min_{i\neq j}\{d_i d_j\}/(2d), 
 \end{equation} 
 where $\bar\delta_m$ is given in Theorem \ref{thm:robust_c}, i.e.,
any trajectories starting within $x(t_o) \in W(\chi)$  
ultimately converge to equilibrium subspace $\mathbb{E}$;
\item[B.] (Angle Perspective) 
 let  
\begin{equation}
\gamma_r=(\pi-\bar\delta_m)\sqrt{\frac{\min_{i\neq j}\{d_i d_j\}}{\max_{i\neq j}\{d_i d_j\}}}, \label{eq:gammar_ce}
\end{equation}
Then, any trajectories starting within $B_c(\gamma)$ for  $\gamma\in[0,\gamma_r]$  ultimately converge to equilibrium subspace  $\mathbb{E}$, i.e., the region of attraction is $\|B_c \t \delta\|\leq \gamma_r$.
\end{itemize} 
 \ethm 
 \proofnow Note that the derivative of $V(\delta)$, along the trajectory of
(\ref{eq:FOV}), is 
\begin{gather*}
\dot{V}\leq-\frac{\lambda_{2}}{n}\kappa(\gamma)\|\delta_{c}\|^{2}.
\end{gather*}
Then, the result follows from Lemma  \ref{lem:flr_e} and Theorem \ref{thm:robust_c}.
\eproof

\bthm \label{thm:robust_le}  Consider microgrid (\ref{eq:FOV}) with $p(t)=0$ and
 energy function (\ref{eq:Vl}).
\begin{itemize} 

\item[A.] (Energy Perspective) $W(\chi)$ is a positively invariant set for any $\chi\in[ 0,f_{r,\max}]$ where 
\begin{equation}
 f_{r,\max}=\gamma^2_{s}\kappa(\gamma_{s})\min_{(i,j)\in\mathcal{E}}\{a_{ij}\}/2, \label{frmax}
 \end{equation} 
where  $\gamma_s$ satisfies (\ref{eq:gamma_sl}),
i.e., any trajectories starting within $x(t_o) \in W(\chi)$ where 
ultimately converge to equilibrium subspace $\mathbb{E}$;
\item[B.] (Angle Perspective) 
let   
\begin{equation}
\gamma_r=\gamma_{s}\sqrt{\kappa(\gamma_s)\frac{\min_{(i,j)\in\mathcal{E}}\{a_{ij}\}}{\max_{(i,j)\in\mathcal{E}}\{a_{ij}\}}}, \label{eq:gammar_le}
\end{equation}
 Then, any trajectories starting within $B_l(\gamma)$ for  $\gamma\in[0,\gamma_r]$ ultimately converge  to equilibrium subspace $\mathbb{E}$,  i.e., the region of attraction is $\|B\t \delta\|\leq \gamma_r$.
\end{itemize} 
 \ethm 
 \proofnow 
  Note that the derivative of $V(\delta)$, along the trajectory of
(\ref{eq:FOV}), is 
\begin{gather}
\dot{V}\leq - \kappa^2(\gamma) \lambda_s(Q) \frac{\min_{(i,j)\in\mathcal{E}}\{a_{ij}\}}{\max_{(i,j)\in\mathcal{E}}\{a_{ij}\}} \|\delta_l\|^{2}
\end{gather}
Then, the result follows from Lemma  \ref{lem:flr_e} and Theorem \ref{thm:robust_l}.
 \eproof
\brem
When a fault occurs, we need to assure that  the power system remains stable  after the fault is cleared. The time duration between the fault occurence and fault clearance is called critical clearing time. 
 If the fault is cleared before the fault-on trajectories reach the boundaries of the region of attraction in Theorem \ref{thm:robust_ce} and \ref{thm:robust_le}, the trajectories can converge back to equilibrium subspace $\mathbb{E}$ again. Therefore, we can use Theorem \ref{thm:robust_ce} and \ref{thm:robust_le} to calcuate the critical clearing time for power systems, which will be demonstrated in Section \ref{sec:sim}.
\erem

\begin{table*}[ht]
\begin{center}
\caption{Transmission line parameters and nominal power profile $p_i^o$\label{tab:line} }
\centering
\begin{tabular}{|c|c|c|c|c|c|c|c|c|c|}
\hline 
line parameters (per unit)& $a_{14}$ & $a_{45}$ & $a_{56}$ & $a_{36}$ & $a_{67}$ & $a_{78}$ & $a_{82}$ & $a_{89}$ & $a_{94}$\tabularnewline
\hline 
$a_{ij}$ (set 1) & 17.2376 & 10.7036 & 5.8484 & 17.1069 & 9.8343 & 13.6459 & 15.8972 & 6.0142 & 11.3837\tabularnewline
\hline 
$a_{ij}$ (set 2) & 8.4148 & 10.6607 & 9.9044 & 10.1356 & 12.2033 & 10.6274 & 13.6683 & 9.5708 & 11.3565\tabularnewline
\hline 
bus number $i$ & 1 & 2 & 3 & 4 & 5 & 6 & 7 & 8 & 9\tabularnewline
\hline 
$\theta^o_i$ (rad, set 1) & 0.1162 & 0.2195 & 0.1406 & 0.0483 & 0.0089 & 0.0909 & 0.0634 & 0.1168 & 0\tabularnewline
\hline 
$\theta^o_i$ (rad, set 2) & 0.1841 & 0.1994 & 0.1269 & 0.0446 & 0 & 0.0429 & 0.0163 & 0.0799 & 0.0009\tabularnewline
\hline 
$p_{i}^{o}$ (per unit) & 1.17 & 1.63 & 0.85 & -0.2 & -0.9 & -0.1 & -1 & -0.2 & -1.25\tabularnewline
\hline 
\end{tabular}%
\end{center}
\end{table*}

\section{Numerical Simulation } \label{sec:sim}
\begin{figure}
\centering
\includegraphics[scale=0.18]{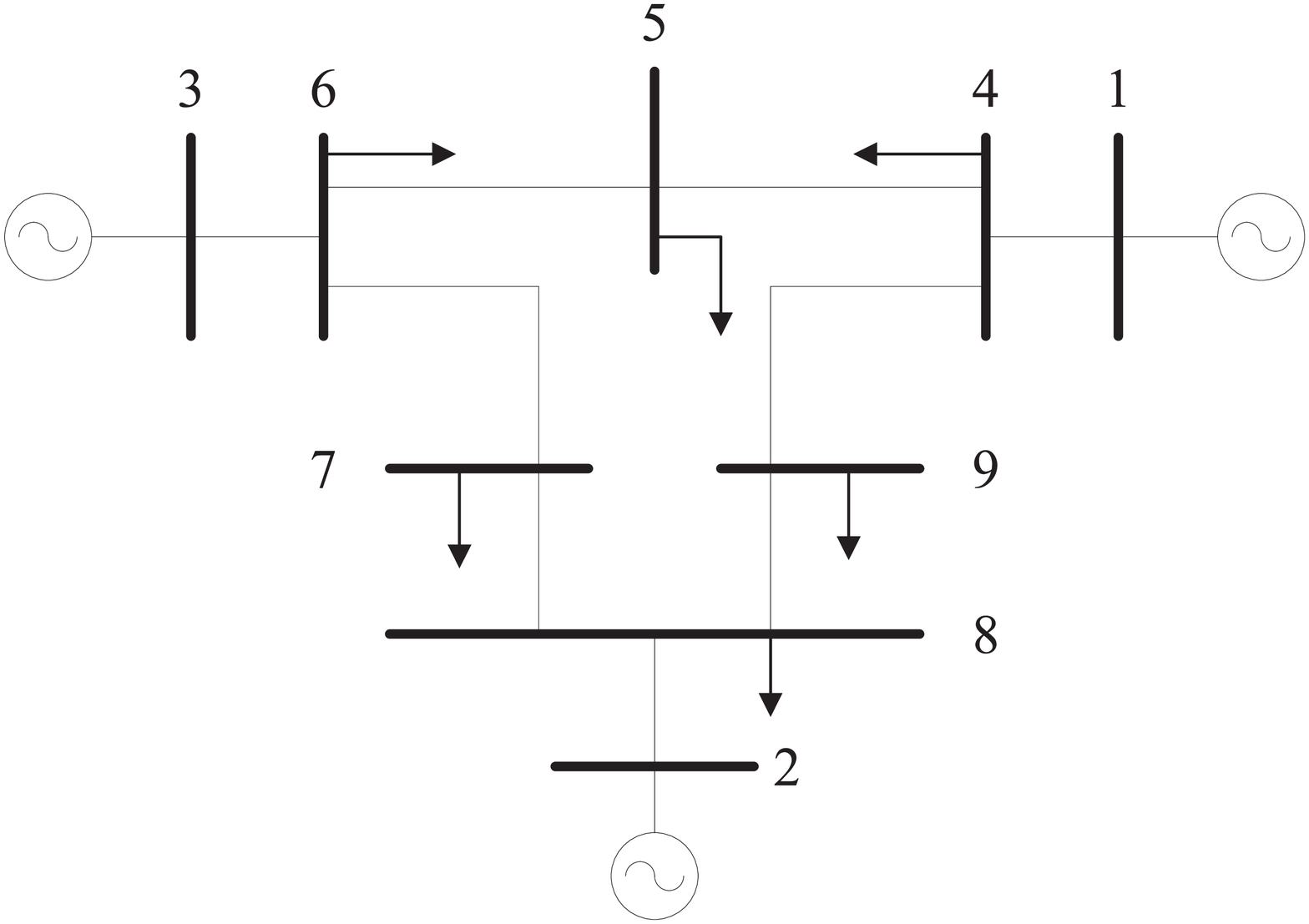}\caption{Microgrids in 9-bus IEEE Test system structure\label{fig:test}}
\end{figure}

Consider lossless microgrids in the network structure of  IEEE 9-bus test system illustrated in
Fig. \ref{fig:test}.  Buses $1,2,3$ are the
inverter-interfaced energy sources while the other buses are load buses. The numerical simulation will compare  two algebraic stability conditions (\ref{eq:sigma_varep}) and (\ref{eq:sigma_varepl}) given in Theorem \ref{thm:robust_c} and \ref{thm:robust_l}, respectively. Both conditions are sufficient conditions for the synchronization. We will use two sets of line parameters $a_{ij}$ and  show that one condition is not necessarily better than the other but they complement each other for predicting the stability and estimating the region of attraction.  Two sets of line parameters $a_{ij}$ and  the nominal power profile $p_o=\mbox{col}\{p_1^o,\cdots,p_n^o\}$ are given in Table. \ref{tab:line}.  Note that $\max_{(i,j)\in\mathcal{E}}\{a_{ij}\}/ \min_{(i,j)\in\mathcal{E}}\{a_{ij}\}=1.6243$ in set 2 compared with $\max_{(i,j)\in\mathcal{E}}\{a_{ij}\}/ \min_{(i,j)\in\mathcal{E}}\{a_{ij}\}=2.9474$ in set 1, showing line parameters in set 2 are more evenly distributed across the network   than   set 1. 
The solutions $\theta_{o}$ to power flow equation (\ref{eq:pf})  can be calculated and illustrated in Table. \ref{tab:line}. As a result, $\bar \theta_l$ in (\ref{eq:theta_l}) is $\bar \theta_{l}=0.2195$ rad  for  set 1 and $\bar \theta_{l}=0.1395$ rad  for  set 2. The system parameter $d_{i}$ is randomly generated within range
$d_{i}\in[0.7,1]$. 

We will emulate two scenarios: the time-varying disturbance scenario with $p\neq0$ and line tripping scenario with $p=0$ in (\ref{eq:FOV}). First, for $t\in[0,5)$s, since the equilibrium point is locally stable, we allow angles to settle
to the equilibrium subspace $\mathbb{E}$. For the time-varying disturbance scenario, after $t=5$s, a random disturbance is injected at bus $1$ to emulate the power generation fluctuation for renewable
power. The disturbance will change its value randomly every $0.1$
s, but its magnitude is bounded, i.e., $\sup_{t\in[0,\infty)}|p(t)|<p_d$ for some constant $p_d$. For line tripping scenario, at $t=5$s, we assume the electric line connecting bus $1$ and $4$ trips causing the system instability and making system trajectory leaves the equilibrium subspace $\mathbb{E}$.  Theorem \ref{thm:robust_ce}  and Theorem \ref{thm:robust_le} give two estimation of region of attraction, namely $\|B_c \delta\|\leq \gamma_r$ for $\gamma_r$ in (\ref{eq:gammar_ce}) and $\|B \delta\|\leq \gamma_r$ for $\gamma_r$ in (\ref{eq:gammar_le}). When  both conditions $C_1:\|B_c \delta\|\geq  \gamma_r$ and $C_2:\|B \delta\|\geq  \gamma_r$ are triggered, we immediately re-close the line and retain the origin system structure making system trajectory converge to the equilibrium subspace $\mathbb{E}$ again. We denote $T_1$ and $T_2$ as the time  when $C_1$ and $C_2$ are triggered, respectively. In fact, $T_1$ and $T_2$  are the critical clearing time based on estimation of region of attraction given by  Theorem \ref{thm:robust_ce}  and Theorem \ref{thm:robust_le}. Note that the larger critical clearing time  is more desirable, which allows more time for the protection system to take actions.

\begin{figure}
\centering
\subfloat[Time-varying disturbance scenario.]{
\includegraphics[scale=0.25]{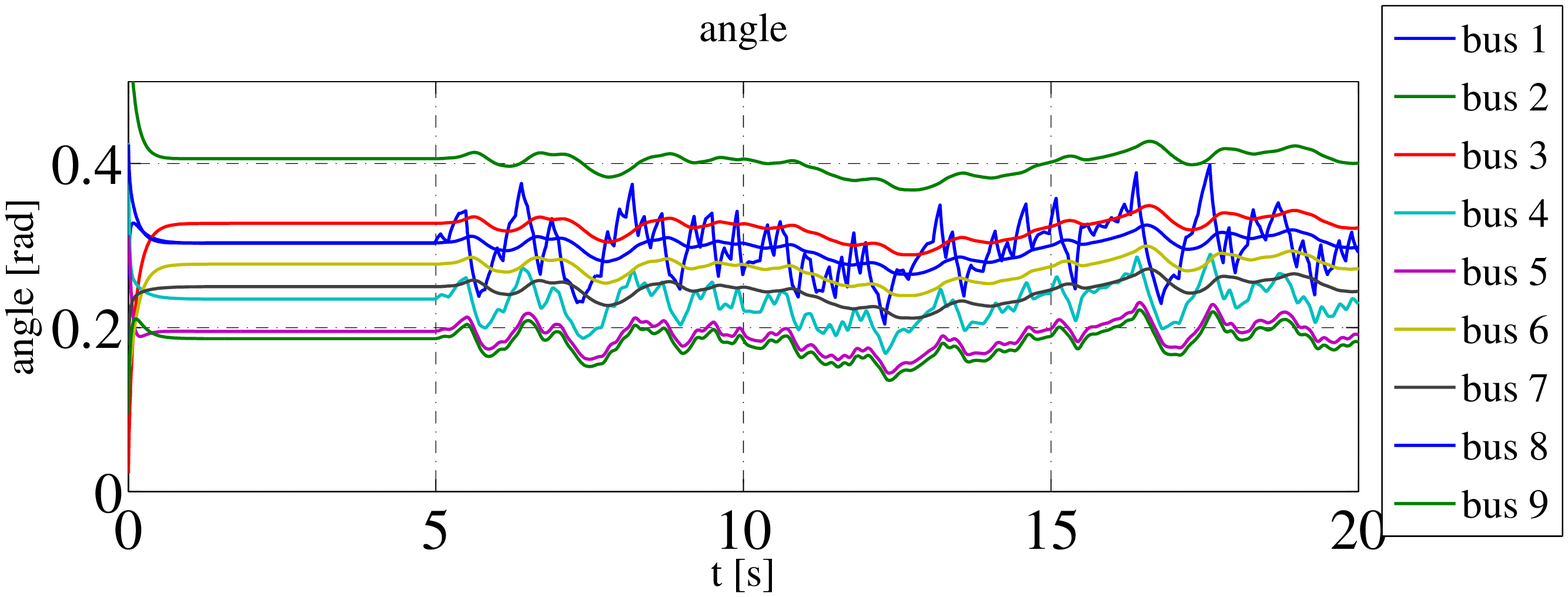}}\\
\subfloat[Line tripping scenario.]{
\includegraphics[scale=0.25]{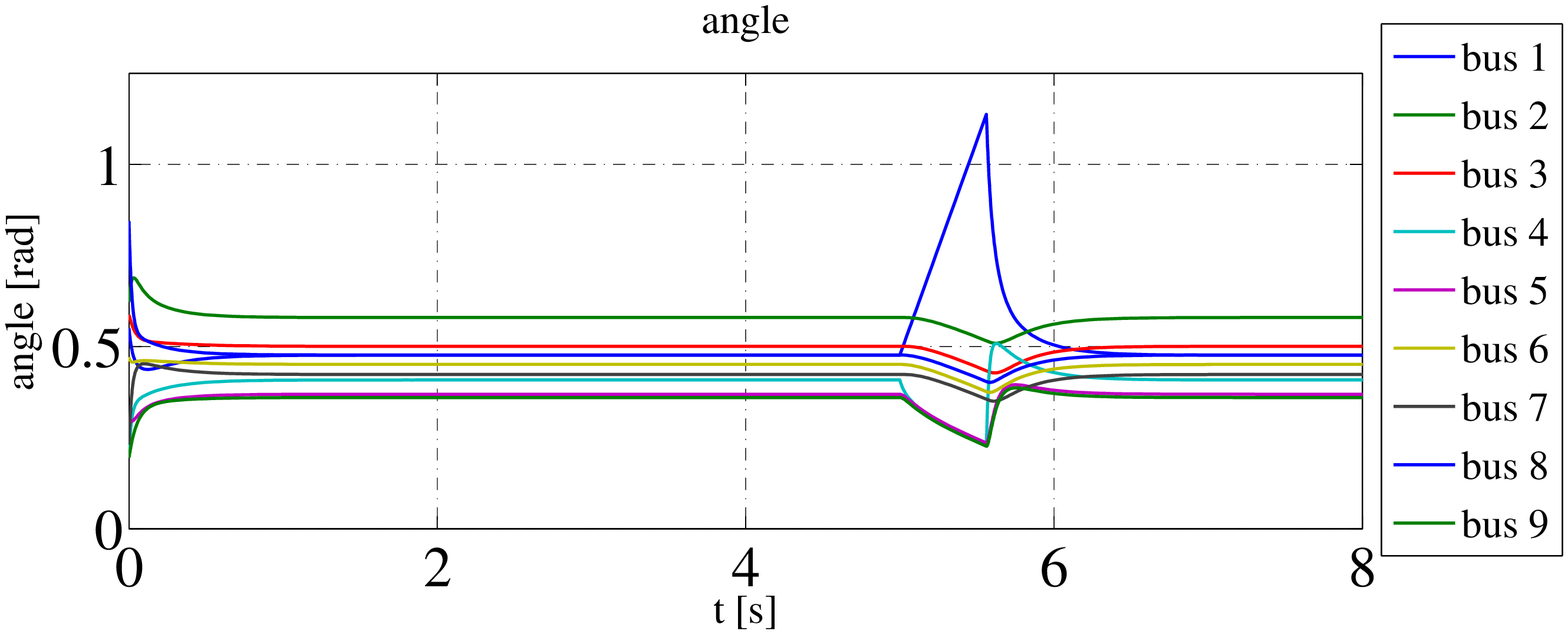}}
\caption{Simulation for parameter set 1.}
\label{fig:sim1}
\end{figure}

For time-varying disturbance scenario with  parameter set 1 , we set disturbance level $p_d=1.0$, one can calculate $\lambda_2=4.0147$ and $\lambda_{cr}=3.961$ such that the algebraic stability condition (\ref{eq:sigma_varep}) is satisfied, while  $\lambda_s(Q)=32.6285$ and $\lambda_{cr}=47.5875$ which shows  the algebraic stability condition (\ref{eq:sigma_varepl}) is not satisfied and hence gives more conservative stability result. For the line tripping scenario, one can calculate  region of attraction  from Theorem \ref{thm:robust_ce} and \ref{thm:robust_le} and obtains two estimations  $C_1:\|B_c \delta\|\leq 2.3044$  and $C_2:\|B \delta\|\leq 0.7115$, respectively.
The numerical simulation shows that the critial clearing time  $T_2=5.4658$s and $T_1=5.6167$s at which the line is re-connected, concluding that the stability result in terms of Definition \ref{def:cohesive_l1} is more conservative than Definition \ref{def:cohesive_c1}. The simulation result is illustrate in Fig. \ref{fig:sim1}.

\begin{figure}
\centering
\subfloat[Time-varying disturbance scenario.]{
\includegraphics[scale=0.25]{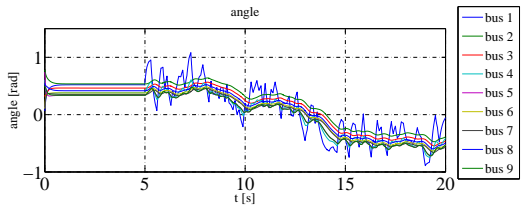}}\\
\subfloat[Line tripping scenario.]{
\includegraphics[scale=0.25]{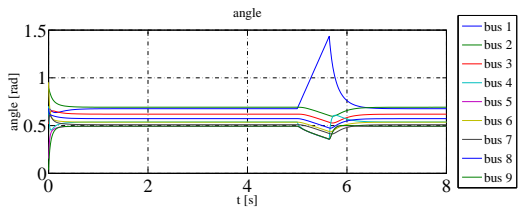}}
\caption{Simulation for parameter set 2.}
\label{fig:sim2}
\end{figure}

For  time-varying disturbance scenario with parameter set 2, we set  disturbance level $p_d=5.0$, one can calculate $\lambda_s(Q)=48.5049$ and $\lambda_{cr}=45.6904$  such that the algebraic stability condition (\ref{eq:sigma_varepl}) is satisfied, while $\lambda_2=4.5773$ and $\lambda_{cr}=45.6904$    which shows  the algebraic stability condition (\ref{eq:sigma_varep}) is not satisfied and hence gives more conservative stability result. For the line tripping scenario, one can calculate the  region of attraction  from Theorem \ref{thm:robust_ce} and \ref{thm:robust_le} and obtains two estimations   $C_1:\|B_c \delta\|\leq 2.2684$  and $C_2:\|B \delta\|\leq 0.9393$, respectively.
The numerical  simulation shows that the critical clearing time $T_1=5.6002$s and $T_2=5.6439$s, concluding that the stability result in terms of Definition \ref{def:cohesive_c1} is more conservative than Definition \ref{def:cohesive_l1}. The simulation result is illustrate in Fig. \ref{fig:sim2}.

\section{Conclusion  } \label{sec:Conclusion}
In this paper, we first presented  the first-order power system model  which coincides with non-uniform Kuramoto oscillators. 
Then, we introduced
two definitions of stability in terms of phase cohesiveness and frequency boundedness.
We proposed the stability analysis framework based on the RPLF and applied it to  derive  stability conditions in terms of two proposed stability definitions. Finally, we explicitly gave the estimation of region of attraction for microgrids. The effectiveness of the
theoretical analysis is verified by the numerical simulation.

\appendix
 \textbf{Proof of Lemma \ref{lem:stab}:}
Since $V(x)$ is a RPLF satisfying (\ref{eq:V_inq}) and (\ref{eq:V_dot_Lemma}),   $V(x)$ decreases outside the ball $B(\mu(\gamma))$.  In fact, $B(\mu(\gamma))$ is contained in $W(\chi)$, 
which follows from   $x\in B(\mu(\gamma)) \Rightarrow x\in W(f_l(\gamma))\Rightarrow x\in W(\chi)$ due to   $\chi\geq f_l(\gamma)$.
Therefore, one has $\dot{V}<0$ at the boundary of $W(\chi)$ and
hence $W(\chi)$ is an invariant set, i.e., any trajectories starting with $x\in W(\chi)$ will converge
to and stay in $W(f_l(\gamma))$. The above argument is based on the argument
$\|x\|\leq\gamma$ (the requirement for the existence of the RPLF). Next, we will show the condition $\|x(t)\|\leq\gamma$ holds for $t\geq t_{o}$ which is true 
if $\chi\leq f_r(\gamma) $, due to 
\begin{gather}
\underline \alpha(\gamma)\|x(t)\|^2\leq V(x(t))\leq\chi\leq f_r(\gamma).\label{eq:V_inq_prf-3}
\end{gather}
Hence, if  $f_r(\gamma)\geq f_l(\gamma)$ or equivalently (\ref{eq:gamma_inq}) holds, it is  guaranteed that the  invariant set $W(\chi)$ within which $\|x(t)\|\leq\gamma$ can be found. The proof is thus complete.
 \eproof

 \textbf{Proof of Lemma \ref{def:sin}:}
 We will prove $\mbox{A} \Rightarrow \mbox{B} \Rightarrow  \mbox{C}$. 
 
$\Rightarrow A$. It is very straightforward to verify Statement A  when $g(\gamma)$ is a quasi-sinusoidal function.  In the rest of the proof, we consider functions $g(\gamma)$, $f_l(\gamma)$ and $f_r(\gamma)$ within the range $\gamma \in [\gamma_{\min},\gamma_{\max}]$.
 
\begin{figure}
\centering
\includegraphics[scale=0.7]{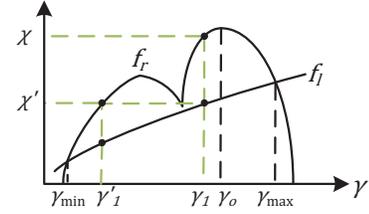}\caption{The relation between $f_l(\gamma)$ and $f_r(\gamma)$.\label{fig:flr}}
\end{figure}

$A \Rightarrow B$.  Statement A  implies that  $ f_l(\gamma) \leq f_r(\gamma)  $ for $\gamma\in[\gamma_{\min},\gamma_{\max}]$,  $ f_l(\gamma_{\min}) = f_r(\gamma_{\min})  $ and $ f_l(\gamma_{\max}) = f_r(\gamma_{\max}) $. By  Lemma \ref{lem:stab},  $\chi\in[f_l(\gamma), f_r(\gamma)]$ whose   maximum range is  $[f_{l,\min},f_{r,\max}]$ as $\gamma$ varies within $\gamma \in [\gamma_{\min},\gamma_{\max}]$. The relation between $f_l(\gamma)$ and $f_r(\gamma)$ is schematically illustrated in Fig. \ref{fig:flr}. For any $\chi\in[f_{l,\min},f_{r,\max}]$,
one can find $\gamma_{1}=\arg\min_{\gamma\in[\gamma_{\min},\gamma_{\max}]}\{f_r(\gamma)=\chi\}$.
Since $f_l(\gamma_1)\leq\chi\leq f_r(\gamma_1)$,
applying Lemma \ref{lem:stab} shows any trajectories starting
within $x\in W(\chi)$
is ultimately contained in the invariant set $W(f_l(\gamma_1))$.
Set $\chi'=f_l(\gamma_1)\geq f_l(\gamma_{\min})$
and repeat the above argument with $\chi$ replaced by $\chi'<\chi$ and find $\gamma'_1$.
It is noted $\chi'$ and $\gamma'_1$  until $\chi'=f_l(\gamma_{\min})$. This search pattern is illustrated in Fig. \ref{fig:flr}.
Then, we can prove the trajectory   is eventually contained in 
set $W( f_l(\gamma_{\min}))$. Statement B is proved.
 
$B \Rightarrow C$. Due to (\ref{eq:V_inq}),    $\bar\alpha(\gamma)\|x\|^2\leq f_{r,{\max}}$ implies $V\leq f_{r,{\max}}$, which shows that $B(\sqrt{f_{r,{\max}}/\bar\alpha(\gamma)}) \in W(f_{r,{\max}})$, i.e., any trajectory starts inside the ball $B(\sqrt{f_{r,{\max}}/\bar\alpha(\gamma)})$ is also inside $W(f_{r,{\max}})$. The size of the ball $B(\sqrt{f_{r,{\max}}/(\bar\alpha(\gamma)})$ depends on the choice of $\gamma$ and is  required to be smaller than $\gamma$ (the requirement of the RPLF for us to use (\ref{eq:V_inq})). We need to seek  such ball of the  largest  size inside  $W(f_{r,{\max}})$, which is equivalent to find $\gamma_{r}$ in (\ref{eq:rls_g}). As a result, $B(\gamma_{r}) \in W(f_{r,{\max}})$.

We can use similar argument to obtain $\gamma_l$.  Because  $V\leq f_{l,\min}$ implies $\underline\alpha(\gamma)\|x\|^2\leq f_{l,\min}$, one has $W(f_{l,\min}) \in B(\sqrt{f_{l,\min})/\underline\alpha(\gamma)})$, i.e., any trajectory entering $W(f_{l,\min})$  also enters  the ball $B(\sqrt{f_{l,\min}/\underline\alpha(\gamma)})$. The size of the ball $B(\sqrt{f_{l,\min})/\underline\alpha(\gamma)})$ depends on the choice of $\gamma$ and is also required to be smaller than $\gamma$. We need to seek  such ball of the  smallest  size that encloses  $W(f_{l,\min})$, which is equivalent to find $\gamma_{l}$ in (\ref{eq:rls_g}). As a result, $W(f_{l,\min}) \in B(\gamma_l) $. Then, the statement C follows from statement B.
 \eproof

\bibliographystyle{plain}
\bibliography{ref}

\end{document}